\documentclass[%
twoside,%
11pt,%
]{memoir}

\usepackage[latin1]{inputenx}
\usepackage[T1]{fontenc}
\usepackage[english]{babel}
\usepackage[square]{natbib}				
\usepackage[dvipsnames]{xcolor} 	
\usepackage{graphicx}			
\usepackage{import} 			
\usepackage{calc}				
\usepackage[open,openlevel=4]{bookmark}
\usepackage{caption,subcaption} 			
\usepackage{microtype}			
\usepackage{pdflscape}			
\usepackage{multicol}			

\graphicspath{{figures/}}

\usepackage{amsmath,amssymb,amsfonts,theorem}
\usepackage{thesis-math}
\usepackage{commath}
\usepackage{bm} 		
\usepackage{gensymb} 	

\usepackage{listings}
\usepackage{algorithm,algpseudocode,etoolbox}
\usepackage{matlab-prettifier}

\usepackage{hyperref}
\colorlet{mylinkcolor}{MidnightBlue}
\colorlet{mycitecolor}{MidnightBlue}
\colorlet{myurlcolor}{MidnightBlue}

\hypersetup{
  linkcolor  = mylinkcolor, 
  citecolor  = mycitecolor,
  urlcolor   = myurlcolor,
  colorlinks = true,
}

\usepackage{fancyhdr}
\fancypagestyle{lscape}{%
\fancyhf{} 
\fancyfoot[LE]{}
\fancyfoot[LO] {}
 
}

\DisemulatePackage{setspace}
\usepackage{setspace}
\midsloppy
\captionnamefont{\footnotesize}
\captiontitlefont{\small}

\setstocksize{297mm}{210mm}
\settrimmedsize{\stockheight}{195mm}{*}
\settypeblocksize{9in}{4.9in}{*}
\setlrmargins{*}{*}{4}
\setulmargins{*}{*}{2}
\setmarginnotes{5mm}{45.23mm}{\onelineskip}
\checkandfixthelayout

\maxsecnumdepth{subsubsection}
\maxtocdepth{subsection}

\makeatletter 
\renewcommand{\fnum@figure}{\textsc{\figurename~\thefigure}}
\makeatother

\makeindex

\headstyles{bringhurst}
\chapterstyle{thatcher}
\maxtocdepth{section}

\begin{document}
\raggedbottom


\thispagestyle{empty}
\phantomsection
\label{titlepage}
\bookmark[dest=titlepage]{Title page}
\calccentering{\unitlength}
\begin{adjustwidth*}{\unitlength}{-\unitlength}
\setlength{\parindent}{0pt}

\centering
\vspace*{5\onelineskip}
{
\Huge
Exponential asymptotics and\\
Stokes surfaces in nonlinear \\
three-dimensional flows \par
}
\mbox{}\rule[0pt]{0.55\textwidth}{1pt} \par  

\vspace*{1\onelineskip}
{\scshape John Fitzgerald} \\
{\itshape MMath Mathematics, University of Oxford}

\vfill 
\includegraphics[width=4cm]{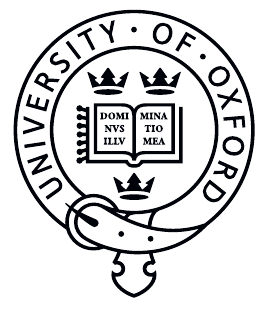}
\vfill
A Part C Dissertation \\

Hilary Term 2018

\end{adjustwidth*}







%
\frontmatter
%
%
%
%
\tableofcontents*
\mainmatter




\chapter{Introduction \label{introduction}} 


%
%

Over the last few years, there has been much development into asymptotic analyses of water waves at low speeds, particularly in connection with wave-structure interactions. As we will see in Sec.~\ref{sec:goveqns}, a crucial nondimensional parameter for such investigations is the Froude number,
\begin{equation} \label{froudenumber}
 	F = \frac{U}{\sqrt{gL}} ,
\end{equation}
which may be thought of as a measure of the ratio between inertial (governed by the typical velocity scale $U$, and length scale $L$) and gravitational (governed by $g$) effects. The principal focus of this essay is to explore the low-Froude limit, $F \to 0$, corresponding to gravity dominating the system. In practice, this is of interest as it provides a means of relating the geometry of a boundary to the generation of free-surface waves.

There is, however, a difficulty introduced by taking this limit -- as was shown by \cite{ogilvie1968wave}, traditional asymptotic methods fail to predict the occurrence of free-surface waves, as they reveal a surface which is flat \emph{beyond-all-orders} of $F$. This is what is now known as the low-speed paradox, the resolution of which requires \emph{exponential asymptotic} techniques, as will be discussed in Sec.~\ref{sec:expasym}. Unfortunately, in resolving one dilemma a peculiar new issue has appeared. It turns out to be the case that these exponentially small terms display what is known as the \emph{Stokes Phenomenon} -- as we traverse the domain, we can observe that they switch-on or -off -- seemingly instantaneously -- across what are analogously called Stokes lines.

\begin{figure}[htb] \centering
	\includegraphics[width=0.9\textwidth]{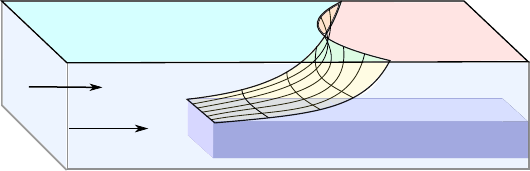}
	\caption{Visualisation of a Stokes surface (shown yellow) coming from flow past a three-dimensional obstruction. Arrows show direction of flow. On the upper surface, blue corresponds to a waveless region, red to the region after intersection with the Stokes surface, with exponentially small waves switched on. \label{3dstokes}}
\end{figure}

In 2D, it is found that these lines emerge from certain points on the boundary of the fluid (or other critical points of the flow) and later intersect the free surface. As we will see in Chap.~\ref{chap:expascrt}, the locations of these lines are related to the solution of an eikonal equation, 
\begin{equation}
 	|\chi(\bm{x})|^{2} = 0 ,
\end{equation}
which, with additional boundary conditions, we may solve on the free surface by Charpit's method (see \emph{e.g.} \cite{ockendon2003applied}). However, to do so we must analytically continue the system, allowing $\bm{x}\in\mathbb{C}$ -- we enter the world of complex ray theory. It is presumed that the comparable situation occurs for three-dimensional flow past an obstacle, where now waves switch-on across a Stokes surface -- the higher-dimensional analogue of Stokes lines. This is visualised in Fig.~\ref{3dstokes}. Note that in actuality the Stokes surface leaving the complexified boundary is now a three-dimensional manifold, so the surface shown would only correspond to the intersection with real space -- the problem of how to determine this intersection is an open question.
\vspace{0.2in}

While a popular current approach to study two-dimensional free-surface flows, reduction of the problem to a one-dimensional boundary integral (\emph{cf.} Sec.~\ref{sec:boundint}), has had great success, it has the drawback of reliance on complex variable methods. As such it has limited extension to three dimensions, where typically we would instead require direct consideration of Laplace's equation for the velocity potential, $\nabla^{2}\phi = 0$. Recently, \cite{lustri2013steady} have managed to apply exponential asymptotics and complex ray theory to do so, for the problem of uniform flow past a point source. This has in fact provided an alternative theory for the well-known case of \cite{kelvin1887ship} waves. The point source potential is significant, as by a superposition of such (\emph{i.e.} by using them as a Green's function, see \emph{e.g.} \cite{havelock1928theory,liu2001transient}) it is possible to model flow past a far more general obstruction. 

\begin{figure}[htb] \centering
	\includegraphics[width=0.8\textwidth]{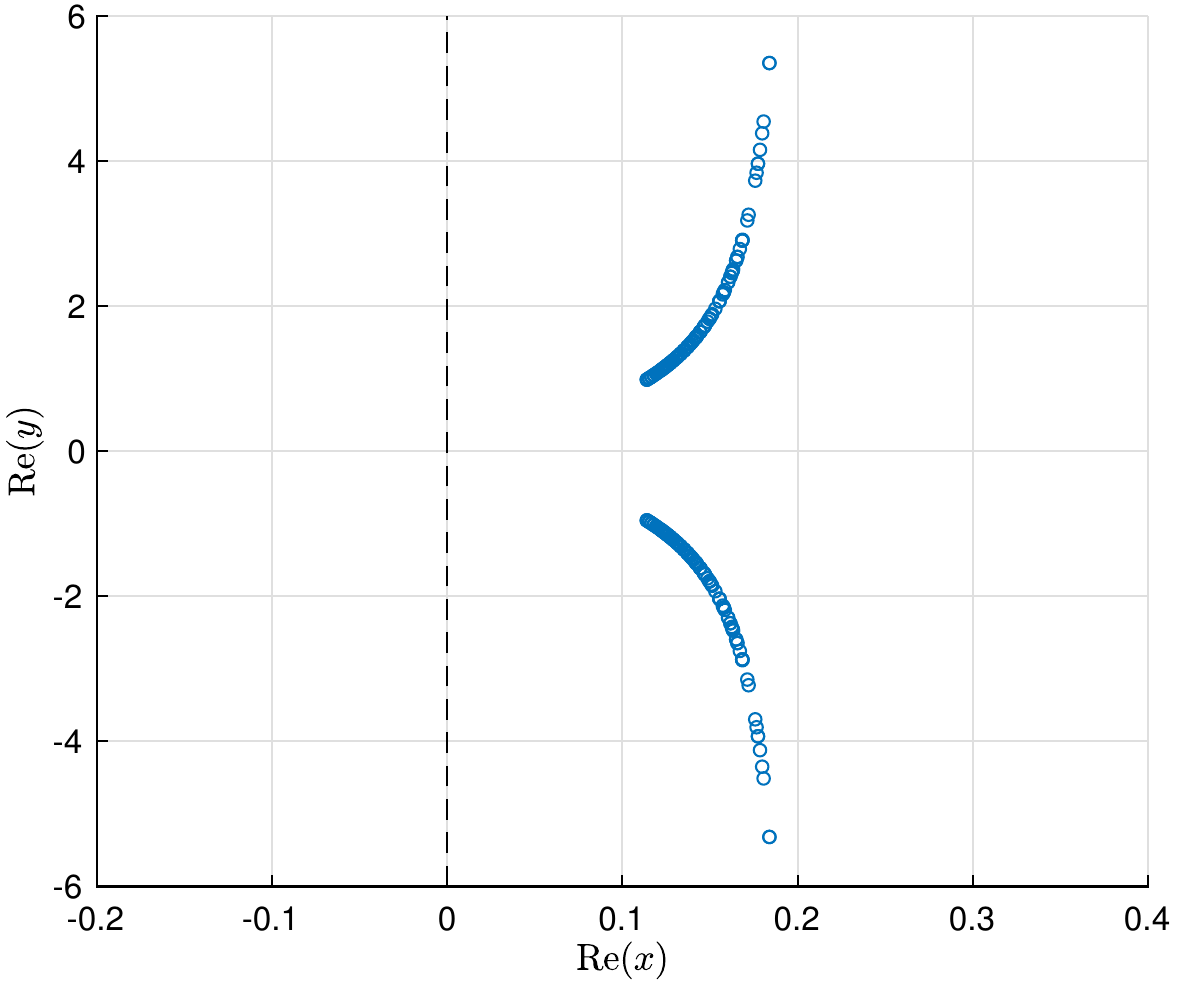}
	\caption{Shown blue are points of intersection (shown blue) of a 4D Stokes surface due to flow past a point source (placed at $(x,\, y, \, z) = (0, \, 0, \, -0.5)$, strength $\delta = -0.3$) with the real free surface. Found with the numerical method of Chap.~\ref{chap:num3d}. Shown dashed black is the intersection predicted by the linear method from \cite{lustri2013steady}. \label{nlsourcewings}}
\end{figure}

However, \cite{lustri2013steady} make the assumption of negligible source strength, $\delta$, so that they may linearise the governing equations and solve for the function $\chi$. In this dissertation, we seek to expand the scope of their method to account for more general flows, and offer an application for the same problem but now with $\delta = \Oh(1)$. We find that in order to solve for the Stokes surfaces of this nonlinear system, we must develop a numerical scheme to shoot from the analytically continued free surface into real space. The principal contribution of this work is shown in Fig.~\ref{nlsourcewings}, displaying the disparity between the line of intersection found by \cite{lustri2013steady} and that found by our own nonlinear numerical method. Our steps are as follows: 

First, in Chap.~\ref{chap:expascrt} we derive the necessary ray equations to deal with nonlinear potentials, and consider necessary initial conditions for application of a numerical scheme.

Next, in Chap.~\ref{chap:raytheoryint} we use the convenience of two dimensions to analytically find Stokes lines in the fluid, for later comparison with numerical results.

Following this, in Chaps.~\ref{chap:num2d}~and~\ref{chap:num3d}, we develop numerical methods to directly find Stokes lines/surfaces for nonlinear two- and three-dimensional flows. For the more difficult 3D situation of Chap.~\ref{chap:num3d} we apply the method for the case of flow past a source, and compare with the results of \cite{lustri2013steady}. 



%

\section{Exponential asymptotics} \label{sec:expasym}

While traditional asymptotic methods provide crucially useful insights into many problems, they face a fatal difficulty: In many physical problems, key features of the solution scale like $\e^{-\text{const.}/\ep}$ where $\ep\ll 1$ is a positive perturbation parameter \citep{boyd1999devil}. As suggested, the problem of free-surface waves for gravity-driven potential flows in the low-Froude limit is one such example. This means that they are \emph{completely invisible} to normal asymptotic expansions in powers of $\ep$. In many practical applications, the perturbation parameter is not sufficiently small to justify ignoring these features, and so a different approach must be taken to produce tolerable accuracy. As Boyd puts it,
\begin{quote}
 	We can no more automatically assume an effect is negligible because it is proportional to $\exp(-q/\ep)$ than a mother can regard her baby as insignificant because it is only sixty centimeters long. \cite{boyd1999devil}
\end{quote}


The field of \emph{exponential asymptotics} has been developed to account for these terms. It began with the idea of \emph{optimal truncation} -- that is (generally, \emph{cf.} \cite{boyd1999devil}) truncating a divergent asymptotic series near its least term -- first used in pursuing solutions of differential equations by \cite{stokes1864discontinuity}. The error in the approximation is typically then exponentially small, and in fact may be reduced yet further -- this process is known as \emph{hyperasymptotics} \citep{berry1990hyperasymptotics}, which involves using a `resurgence formula' obtained by \cite{dingle1973asymptotic} to sequentially form an asymptotic series for the remainder. This process terminates, and the resulting error is generally less than the square of that found by optimal truncation -- a considerable improvement! The key issue is that these exponentially small terms display the Stokes Phenomenon previously discussed.

\begin{figure}[htb] \centering
	\includegraphics[width=0.8\textwidth]{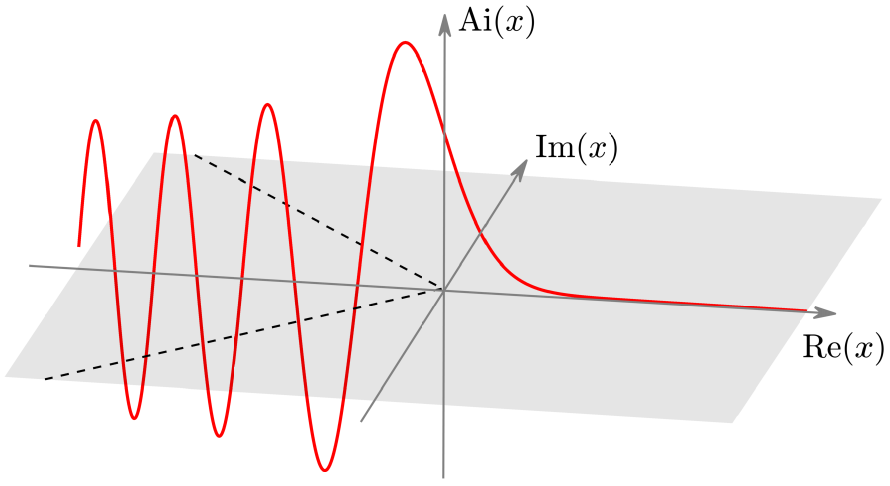}
	\caption{The Airy function of the first kind, $\mathrm{Ai}(x)$, shown red. The complex $x$-plane is shown grey, and Stokes lines dashed black. \label{airywithstokes}}
\end{figure}

A classic example of this is the Airy function of the first kind, $\mathrm{Ai}(x)$, for which we may obtain (from \emph{e.g.} the WKB method \citep{howison2005practical} or the method of steepest descents \citep{bleistein1975asymptotic}) the leading-order asymptotic approximations as $|x| \to \infty$
\begin{equation} \label{Airyform}
	\Ai(x) \sim 
	\begin{cases}
		\frac{x^{-\frac{1}{4}}}{2\sqrt{\pi}}\e^{-\frac{2}{3}x^{\frac{3}{2}}}  & x > 0,  \\
		\frac{x^{-\frac{1}{4}}}{2\sqrt{\pi}}\left(\e^{-\frac{2}{3}x^{\frac{3}{2}}} - \e^{\frac{2}{3}x^{\frac{3}{2}}-\frac{\im\pi}{2}}\right)  \qquad & x < 0.
	\end{cases}
\end{equation}
We see that as $x$ crosses zero, a switching occurs -- the Stokes Phenomenon in action. If we analytically continue the function, allowing $x \in \mathbb{C}$, with the key conditions introduced in the subsequent chapter, \eqref{eq:dingle}, it is seen that this occurs across the Stokes lines $\mathrm{Arg}(x) = \pm 2\pi/3$. Fig.~\ref{airywithstokes} shows $\mathrm{Ai}(x)$ for real argument, along with the Stokes lines in the complex $x$-plane. 
%


\section{Statement of originality}
The nonlinear ray equations derived in Chap.~\ref{chap:expascrt}, and discussion of necessary initial conditions for a numerical scheme therein, are original work. The linearised source problem of the same chapter follows from the work of \cite{lustri2013steady}, and is considered for two reasons: First, it provides us the means to test a 3D numerical method in Sec.~\ref{sec:numlinsource}. It also allows comparison with the results from the nonlinear method of Sec.~\ref{sec:numnlsource}. The integral approach of Chap.~\ref{chap:raytheoryint} finds analogues in other papers, \emph{e.g.} \cite{chapman2006exponential}, and is predominantly included so as to provide figures for later comparison with the numerical results of Chap.~\ref{chap:num2d}. The work of both subsequent chapters is original -- the principal contribution of this dissertation is found in Chap.~\ref{chap:num3d}, where we produce a method to solve numerically for Stokes surfaces, and provide an example application to flow past a nonlinear source.

\chapter{Formulation of the exponential asymptotics and complex-ray problem in the low-Froude limit \label{chap:expascrt}}

As we briefly mentioned in Chap.~\ref{introduction}, the mysterious switching-on or -off of exponentially small terms in the asymptotic expansion of a function, as first considered by G. G. Stokes (see \emph{e.g.} \cite{stokes1864discontinuity}), is known today as the Stokes Phenomenon. As we noted (and Stokes himself believed), at first glance this switching seems instantaneous, but more modern techniques in exponential asymptotics have shown that in fact it occurs smoothly, within a boundary layer region that shrinks to zero as the small parameter $\epsilon \to 0$ \citep{berry1989uniform}. To reiterate, the lines (or in higher dimensions, surfaces) across which this switching occurs are known as Stokes lines (or surfaces). 

Generally, the locations of Stokes lines are predicted using a set of universal rules applicable to a range of linear and nonlinear differential equations. One such rule, as specified in the work of \cite{dingle1973asymptotic}, is as follows. Consider a differential equation for $f = f(z)$ where in a given location of the complex plane, $z\in\mathbb{C}$, 
\begin{equation}
	f(z) \sim f_1(z) = A_1(z) \e^{u_1(z)/\epsilon},
\end{equation}
as $\epsilon \to 0$. Such might be derived using a standard WKB analysis. 
We also assume that there exists a subdominant exponential form given by 
\begin{equation}
	f_2(z) = A_2(z) \e^{u_2(z)/\epsilon},
\end{equation}
which is initially not present. 

The criterion according to \cite{dingle1973asymptotic} is that as $f_1(z)$ is analytically continued across a Stokes line, the subdominant exponential will switch-on, which we write as 
\begin{equation}
	f_1(z) \mapsto f_1(z) + f_2(z).
\end{equation}
This Stokes line occurs where (i) the term being switched-on is of equal phase to the base term present; and (ii) is maximally or exponentially subdominant. That is, 
\begin{equation} \label{eq:dingle}
 	\Im u_{1} = \Im u_{2}, \quad \Re u_{1} \geq \Re u_{2}.
\end{equation}
The \cite{dingle1973asymptotic} criterion above can be thought of as analogous to the sudden switching-on of saddle contributions when using the method of steepest descents -- indeed there is often correspondence between this and the ray theory methods we shall use (see \cite{chapman1999theory},~Sec.~{2.1}, and \cite{trinh2017reduced} for an explicit example in the case of free-surface waves). As we will see in Chap.~\ref{chap:num2d}, the condition of equal phase in particular will be a key property for use in our numerical method.

\section{Governing equations for free-surface flows \label{sec:goveqns}}

Throughout this dissertation we will be focusing on steady, irrotational, free-surface flow of an incompressible, inviscid fluid under gravity. Further, we will ignore the effect of surface tension, and nondimensionalise with respect to a typical length-scale, $L$, and typical velocity-scale, $U$, such that
\begin{equation}
	\bm{x} = L\hat{\bm{x}}, \quad \bm{u} = U\hat{\bm{u}} ,
\end{equation}
where $\bm{u}(\bm{x})$ denotes the flow velocity at the point $\bm{x}$ in the fluid. As we are considering irrotational flow, there exists a \emph{velocity potential}, $\phi$, such that 
\begin{equation}
 	\bm{u} = \nabla\phi = U\hat{\nabla}\hat{\phi} ,
\end{equation}
where the nondimensional velocity potential satisfies $\phi = UL\hat{\phi}$. Immediately dropping the hats, as derived in \emph{e.g.} \cite{acheson1990elementary} we then have the dimensionless governing equations
\begin{subequations} \label{chap1:nondim}
\begin{align}
 	\nabla^{2}\phi &= 0 , && \text{in fluid} , \label{chap1:laplace} \\
	\frac{\ep}{2}(|\nabla\phi|^{2} - 1) + z & = 0 , && \text{on free surface} , \label{chap1:nondimbern} \\
	\nabla\phi\cdot\bm{n} & = 0 , && \text{at boundaries} .
\end{align}
\end{subequations}
Here, $\bm{n}$ denotes the unit normal to the boundary, pointing out of the fluid, and the important nondimensional parameter, $\ep$, denotes the square of the Froude number given by \eqref{froudenumber}, \emph{i.e.}
\begin{equation}
 	\ep = \frac{U^{2}}{gL} .
\end{equation}
As such, the low-Froude limit corresponds to $\ep \to 0$. 

Note that in \eqref{chap1:nondimbern} there is a choice of constant. Within this dissertation, in each problem considered we impose a uniform flow in the $x$-direction of speed $U$ far upstream. As such, after nondimensionalisation we have $\nabla\phi \to \bm{e}_{x}$ as $x\to -\infty$, where $\bm{e}_{x}$ denotes the unit vector in the $x$-direction. Applying a radiation condition that there are no waves as $x \to -\infty$, we set $z \to 0$ on the free surface by taking the given constant. 

\section{Boundary-integral methods and their limitations \label{sec:boundint}}

In general, the system~\eqref{chap1:nondim} is difficult to solve, but there are a variety of means available to proceed. If the problem is two-dimensional, 
one such method is converting the problem into a complex one-dimensional boundary integral. 

Considering such 2D flow, for a point $z = x + \im y$ in the fluid we introduce the complex potential
\begin{equation} \label{comppot}
 	w(z) = \phi(z) + \im\psi(z) , 
\end{equation}
where $\phi$ is the velocity potential as before, and $\psi$ is the streamfunction -- constant on streamlines of the flow. We then have that
\begin{equation} \label{qtheta1}
 	\dd{w}{z} = u - \im v = q\e^{-\im\theta} ,
\end{equation}
where $q = |\mathbf{u}|$ is the flow speed and $\theta$ is the angle the streamline makes with the positive $x$-axis. We also introduce a conformal map, $\zeta$, which takes the fluid domain in the $w$-plane to the UHP -- this will depend on the type of problem approached. This use of complex variable techniques restricts extension to 3D -- in particular, the method cannot be used to determine the location at which exponentially small terms first switch-on. As such, it is not of especial import for the aims of this dissertation, so for brevity we will not elaborate further. The governing equations are then \citep{vanden2010gravity,trinh2017reduced}
\begin{subequations} \label{chap1:boundintsys}
\begin{align}
	\ep\zeta q^{2}\dd{q}{\zeta} &= \sin\theta \\
 	\log q(\zeta) \mp \im\theta(\zeta) &=  \mathrm{BI}(\zeta) -\frac{1}{\pi} \int_{0}^\infty \frac{\theta(\zeta')}{\zeta' - \zeta} \, \de{\zeta'} ,
\end{align}
\end{subequations}
where the minus/plus corresponds to $\zeta$ in the UHP/LHP respectively, and 
\begin{equation}
 	 \mathrm{BI}(\zeta) =  -\frac{1}{\pi}\int_{-\infty}^{0} \frac{\theta(\zeta')}{\zeta' - \zeta} \, \de{\zeta'} ,
\end{equation}
depends solely on the geometry of the boundary.

We will use these equations in Chap.~\ref{chap:raytheoryint} to find leading-order behaviour for the problem of flow over a rectangular step. However, to allow consideration of three-dimensional flows, our principal approach will instead proceed directly from the governing equations \eqref{chap1:nondim}.

\section{The 2D ray equations \label{sec:2drayeqns}}
\subsection{Derivation of the ray equations}

Another common approach to the system \eqref{chap1:nondim} is the assumption that the disturbance of the flow is small, so the equations may be linearised about the free stream (as performed in \cite{lustri2013steady}). Here we instead seek to find a means of considering the full system, initially in 2D before further extension in Sec.~\ref{sec:3drayeqns}. The first step we take is to change coordinates such that
\begin{equation}
 	Z = z - \eta(x), \quad \varphi(x, Z) = \phi(x, z) ,
\end{equation}
where $z = \eta(x)$ corresponds to the free surface. Applying the chain rule (with primes here denoting differentiation with respect to $x$), we find \eqref{chap1:nondim} transforms to 
\begin{subequations}
\begin{align}
 	\varphi_{xx} + \varphi_{ZZ} + \eta'^2 \varphi_{ZZ} 
	- \eta'' \varphi_Z - 2 \eta' \varphi_{xZ} &= 0 && \text{in fluid}, \label{nl2dlaplace} \\
	\varphi_Z(1 + \eta'^2) &= \eta' \varphi_x && Z = 0, \label{nl2dkbc} \\
	\frac{\epsilon}{2} \left[ \varphi_x^2 + (1 + \eta'^{2})\varphi_Z^2 - 2\eta'\varphi_{x}\varphi_Z - 1 \right] + \eta &= 0 && Z = 0. \label{nl2ddbc}
\end{align}
\end{subequations}
Note that there is also a transformed kinematic condition, which must apply at solid boundaries away from the free-surface. However, for reasons that will become clear, this will not explicitly feature in subsequent analysis. As we are interested in the limit $\ep\to 0$, we now try conventional asymptotic expansions for $\varphi$ and $\eta$ as power series in $\ep$, that is 
\begin{equation} \label{phietapower}
 	\varphi = \sum_{n=0}^\infty \epsilon^n \varphi_n \quad \text{and} \quad \eta = \sum_{n=0}^\infty \epsilon^n \eta_n.
\end{equation}
We quickly observe that \eqref{nl2ddbc} then provides $\eta_{0} = 0$, and that using this in \eqref{nl2dkbc} we find $\varphi_{0 Z}(x,0) = 0$. However, further terms are difficult to obtain and as $n$ increases, the $\Oh(\epsilon^n)$ expressions become increasingly convoluted. But mercifully there is a way we may proceed, the key being that we only require the dominant terms in the limit $n \to \infty$ in order to deduce $\chi$. These ideas of estimating the divergent tails of asymptotic approximations follow from the work \cite{chapman1998exponential} (for ODEs) and \cite{chapman2005exponential} (for PDEs). The key ideas from the methodology are summarised as follows:
\begin{enumerate}[(i)]
	\item For singularly perturbed differential equations where a small parameter multiplies the highest derivative, as in the algebraic expansion \eqref{nl2ddbc}, we have a relationship between higher-order terms and the derivatives of lower-order terms. This means that if there is a singularity in the lower-order terms, differentiation will generically increase the power of the singularity and cause the expansion to diverge like a factorial-over-power. 

	\item This idea is analogous to Darboux's Theorem \citep{darboux1878approximation} that the late-order terms of a Taylor series are dominated by the behaviour of the nearest singularity to the point of expansion of the series. This leads to suggestion of the ansatz
	\begin{equation} \label{eq:fnlate}
	 	f_{n} \sim \frac{F\Gamma(n+\gamma)}{\chi^{n+\gamma}} ,
	\end{equation}
	where $\Gamma$ denotes the Gamma function, and $F, \, \gamma$ and $\chi$ are functions that do not depend on $n$. The function $\chi$ is known as the \emph{singulant}, and is zero at singularities of the leading-order solution. The approximation for $f$ at a given point is then formed by summing over all active such terms. Important to note is that in using this ansatz we are making the assumption that the singularities are well-separated -- as shown in \cite{trinh2015exponential}, in problems with coalescing singularities it is necessary to use a more general exponential-over-power form. 
	
	\item As used in \cite{trinh2013new}, the optimal truncation point, $N$, of a series is typically where adjacent terms are approximately equal, \emph{i.e.} where
	\begin{equation}
	 	\left\lvert\frac{\ep^{N}f_{N}}{\ep^{N - 1}f_{N - 1}}\right\rvert \sim 1 .
	\end{equation}
	Under the ansatz above, we see that $N \sim |\chi|/\ep$, and so for any fixed point with $\chi \neq 0$, $N \to \infty$ as $\ep \to 0$. As a result, it is precisely the behaviour of these late-order terms that governs the Stokes switching which we wish to determine. 
	
	\item At optimal truncation, it can be shown that the remainder is expressed in terms of a WKB-type ansatz, 
	\begin{equation} \label{eq:termswitched}
	 	f \sim A\e^{-\chi/\ep} ,
	\end{equation}
	and somewhat coincidentally, this establishes a connection between the exponentially-small remainder and the late-order ansatz \eqref{eq:fnlate} (see the $\chi$ in both).

	Thus we may use the previous condition \eqref{eq:dingle} as providing the location of Stokes lines (or surfaces). As the outer solution (that found by traditional series expansion in powers of $\epsilon$) is only algebraically dependent on $\ep$, this provides the conditions
	\begin{equation} \label{dinglecdn}
	 	\Im(\chi) = 0 , \quad \Re(\chi) \ge 0 .
	\end{equation}
	
	\item The singularity which produces the largest switching term is that which leads to the smallest value of $|\Im(\chi)|$ on the real axis. Typically this is the nearest singularity to the real axis.

\end{enumerate}

Briefly, we note here that in exponential asymptotics problems for nonlinear PDEs, there are a variety of complicated (and not particularly well-understood) effects that may arise; these include the generation of previously unpredicted singularities, and also the so-called \emph{second-generation} Stokes Phenomenon. We do not believe these effects will significantly change the results we present in this work, and we shall omit further discussion of these types of effects (see \emph{e.g.} \cite{howls2004higher,chapman2005exponential}).

Returning to our power series, \eqref{phietapower}, we now make use of the above ansatz so that
\begin{equation} \label{loansP}
	\varphi_n \sim \frac{A(x, Z)\Gamma(n+\gamma)}{\chi(x, Z)^{n+\gamma}} \quad \text{and} \quad
	\eta_n \sim \frac{B(x, Z)\Gamma(n+\gamma)}{\chi(x, Z)^{n+\gamma}} .
\end{equation}
We will assume immediately that $\gamma$ is constant (as in other papers, \emph{e.g.} \cite{chapman2006exponential}), but this could be shown later by looking at lower-order terms in $n$. The determination of the dominant terms below follow from similar ideas from related works of \cite{chapman2006exponential,trinh2011waveless,lustri2013steady}. 

First, in the limit $n\to\infty$, higher indices are dominant over lower indices, \emph{e.g.} $\varphi_{n+1} \gg \varphi_n$. For example
\begin{equation}
	\frac{\varphi_{n+1}}{\varphi_n} \sim \frac{\Gamma(n+1+\gamma)}{\chi \Gamma(n+\gamma)} \sim \frac{n+\gamma}{\chi} = \Oh(n).
\end{equation}
Secondly, differentiation of a term will increase the order of magnitude in $n$ by one. For example, 
\begin{equation} \label{lodifforders}
	\frac{\varphi_{n x}}{\varphi_n} \sim \frac{A_{x}}{A} - \frac{(n+\gamma)\chi_{x}}{\chi} = \Oh(n).
\end{equation}
Hence informally to extract the relevant terms at $O(\epsilon^n)$, we can show that the following quantities are of equivalent order: 
\begin{equation}
	\varphi_n \sim \varphi_{(n-1)x} \sim \varphi_{(n-2)xx} \sim \varphi_{(n-1)Z} \sim \ldots
\end{equation}
Thus at $\Oh(\epsilon^n)$ of \eqref{nl2dlaplace}, \eqref{nl2dkbc} as $n\to \infty$ we find
\begin{gather}
 	\Bigl[\underbrace{\varphi_{n xx} + \varphi_{n ZZ}}_{\text{leading order}} \Bigr] - 2 \underbrace{\eta_{1}'\varphi_{(n-1) xZ}}_{\text{second order}} + \ldots = 0 , \\
	\Bigl[\underbrace{\varphi_{n Z} - \eta_{n}'\varphi_{0 x}}_{\text{leading order}} \Bigr] - \underbrace{\eta_{n-1}'\varphi_{1 x}}_{\text{second order}} + \ldots = 0, 	
\end{gather}
and finally the $\Oh(\epsilon^n)$ terms of \eqref{nl2ddbc} yield
\begin{multline}
	\Bigl[\underbrace{\varphi_{0 x}\varphi_{(n-1) x} + \eta_{n}}_{\text{leading order}} \Bigr] \\ + \Bigl[ \underbrace{\varphi_{1 x}\varphi_{(n-2) x} + \varphi_{1 Z}\varphi_{(n-2) Z} - 2\eta_{1}'\varphi_{(n-2) Z}}_{\text{second order}}\Bigr] + \ldots = 0 .
\end{multline}
It should be noted that as we can see from \eqref{lodifforders}, differentiated expressions contain other lower-order terms. Those stemming from the `leading order' braces would be needed, in addition to the largest magnitude terms of the `second order' expressions, to develop a system of transport equations -- required for calculation of the prefactors $A, \, B$. For the sake of brevity we will not do so here (\emph{cf.} \cite{lustri2013steady},~Sec.~{3.2} for an idea of the method).


Hence, substituting in the ansatz \eqref{loansP}, we find to leading order
\begin{subequations}
\begin{align}
	\chi_x^2 + \chi_Z^2 &= 0 , \\
	A\chi_Z - B\chi_x \varphi_{0x} &=0 , \\
	-A \chi_x \varphi_{0x} + B &= 0.
\end{align}
\end{subequations}
Solving the final two equations for non-trivial $\chi$ requires $\chi_{Z} = (\chi_{x}\varphi_{0x})^{2}$, and thus from the first equation we have (on $Z=0$)
%
\begin{equation} 
	\chi_x^2 + (\chi_x \varphi_{0x})^4 = 0 .
\end{equation}
Note that from the chain rule for the change of coordinates, we have
\begin{equation}
 	\varphi_{0 x}(x,Z) = \phi_{0 x}(x,z) + \eta_0'\phi_{0 z}(x,z).
\end{equation}
However, since $\eta_{0} = 0$, to leading order we may use $\phi_{0 x}(x,0)$ instead of $\varphi$. Thus 
\begin{equation} \label{2dnonlineikonal}
 	\chi_x^2 + (\chi_x \phi_{0x})^4 = 0 .
\end{equation}
To form ray equations, we may apply Charpit's method -- letting $p = \chi_{x}$, and $\tau$ be the characteristic variable along the rays, we have
\begin{equation} \label{2dcharpiteqns}
 	\dd{x}{\tau} = 2p + 4p^{3}\phi_{0 x}^{4} , \quad  \dd{p}{\tau} = -4p^{4}\phi_{0 x}^{3}\phi_{0 xx} , \quad \dd{\chi}{\tau} = 2p^{2} + 4p^{4}\phi_{0 x}^{4}.
\end{equation}
If we now use \eqref{2dnonlineikonal}, the final equation becomes
\begin{equation}
 	\dd{\chi}{\tau} = -2p^{2}.
\end{equation}
It is important to note that in our problem, we would not expect singularities in the leading-order velocity potential on the real free-surface itself \citep{lustri2013steady}. The ray equations must be solved as an initial-value problem tracked from the singularities of $\chi$ (this will depend on the chosen geometry, and will be clarified later). That is, we must appeal to \emph{complex ray theory}. We apply the same ray equations but analytically continue the system into complex space. This allows all of our variables to be complex, and hence this is considered a complexification of the free-surface. Note that in some situations (for instance the boundary-integral method discussed in Sec.~\ref{sec:boundint}), the process of analytic continuation may not be quite as simple. For an introduction to complex ray theory, and some of the potential applications (and issues) therein, we recommend \cite{lawry1996complex, chapman1999theory,kravtsov2005geometrical}.

\subsection{Initial conditions \label{sec:2dic}}

Our procedure below is a general one, and so we have not yet considered the form of the geometry, as specified by $\phi_{0}$. We shall do this in Sec.~\ref{sec:initsource}, but the reader should keep in mind that a given geometry is typically associated with a given singularity structure for the analytic continuation of the leading-order solution, $\phi_0$. 

Consider the initial conditions that would be required by a numerical integration procedure. Intuitively, we may attempt to begin integration of the ray equations exactly from the singularity, where $\phi_{0} \to \infty$, at $x = x_{0}$ say, and by definition $\chi = 0$. 
Unfortunately, the dependence of \eqref{2dcharpiteqns} on derivatives of $\phi_{0}$ presents a dilemma -- at the singularity itself these are infinite and so our ray equations are non-invertible. Instead, we must start slightly away from the singularity, and make the assumption that solutions are sufficiently well-behaved (\emph{i.e.} integrable) such that we may still apply $\chi(x_{0}) \approx 0$. That is we have 
\begin{equation} \label{2dcharpic}
 	x_{0, \rho, \nu} = x_{0} + \rho\e^{\im\nu} , \qquad \chi_{0} = \tilde{\delta},
\end{equation}
with $0<\rho\ll 1$, $0\le \nu < 2\pi$, and $\tilde{\delta} \to 0$ as $\rho \to 0$. Note, in the application of our numerical methods in Chap.~\ref{chap:num2d} we will directly take $\tilde{\delta} = 0$. For the initial condition $p=p_{0}$, \eqref{2dnonlineikonal} provides
\begin{equation}
 	p_{0}^{2}+(p_{0}\phi_{0 x})^{4} = 0 ,
\end{equation}
and so either $p_{0} = 0$ or 
\begin{equation} \label{2dp0ic}
 	p_{0} = \pm\frac{\im}{\phi_{0 x}^{2}(x_{0},0)} .
\end{equation}
Were we to choose $p_{0} = 0$, under our assumption of integrability away from the singularity this leads to a trivial $\chi$ solution. Hence we have that \eqref{2dp0ic} is the correct condition to take, evaluated at $x_{0, \rho, \nu}$. This ends our current development of the ray equations for the 2D problem, and we now perform the analogous analysis for the 3D problem.

\section{The 3D ray equations \label{sec:3drayeqns}}

Now in 3D, we perform the same procedure of exponential asymptotics that converts the governing equations \eqref{chap1:nondim} to a ray-theoretic framework. We now have the two-dimensional free surface $\eta = \eta(x,y)$, and under the transformation 
\begin{equation}
	Z = z - \eta , \qquad \Phi(x,y,Z) = \phi(x,y,z) ,
\end{equation}
with $\nabla = (\partial_{x}, \partial_{y}, \partial_{Z})$, the system becomes 
\begin{subequations}
\begin{align}
	\nabla^{2}\Phi + (\nabla\eta\cdot\nabla\eta)\Phi_{ZZ} - 2\nabla\eta\cdot\nabla(\Phi_{Z}) - \Phi_{Z}\nabla^{2}\eta & = 0 , \label{nl3dlaplace} \\
	\frac{\ep}{2}((\nabla\Phi\cdot\nabla\Phi) + (\nabla\eta\cdot\nabla\eta)\Phi_{Z}^{2} - 2\Phi_{Z}\nabla\eta\cdot\nabla\Phi) + \eta & = 0 , \label{nl3ddbc} \\
	\nabla\eta\cdot\nabla\Phi - (1 + (\nabla\eta\cdot\nabla\eta))\Phi_{Z}& = 0 . \label{nl3dkbc} 
\end{align}
\end{subequations}
Expanding $\Phi, \, \eta$ in linear powers of $\epsilon$
%
\begin{equation} \label{3dloansatz}
 	\Phi = \sum_{n=0}^\infty \epsilon^n \Phi_n \quad \text{and} \quad \eta = \sum_{n=0}^\infty \epsilon^n \eta_n ,
\end{equation}
we see that \eqref{nl3ddbc} provides $\eta_{0} = 0$ once more. This in turn means that \eqref{nl3dkbc} provides $\Phi_{0 Z}(x,y,0) = 0$. To progress, we apply the factorial-over-power ansatz
\begin{equation} \label{loans3d}
	\Phi_n \sim \frac{A(x, y, Z)\Gamma(n+\gamma)}{\chi(x, y, Z)^{n+\gamma}} \quad \text{and} \quad
	\eta_n \sim \frac{B(x, y, Z)\Gamma(n+\gamma)}{\chi(x, y, Z)^{n+\gamma}},
\end{equation}
to describe the late terms as $n\to\infty$. Following this through, we find the leading-order equations as given by
\begin{subequations}
 	\begin{align}
	 	\Phi_{n xx} + \Phi_{n yy} + \Phi_{n ZZ} + \cdots &= 0 , \\
		\Phi_{0 x}\Phi_{(n-1) x} + \Phi_{0 y}\Phi_{(n-1) y} + \eta_{n} + \cdots &= 0 , \\
		\eta_{n x}\Phi_{0 x} + \eta_{n y}\Phi_{0 y} - \Phi_{n Z} + \cdots &= 0.
	\end{align}
\end{subequations}
Hence substituting the ansatz \eqref{loans3d} and simplifying yields the leading-order system
\begin{subequations}
 	\begin{align}
	 	\chi_{x}^{2} + \chi_{y}^{2} + \chi_{Z}^{2} &= 0 , \label{eikonal3d1}\\
		-A(\chi_{x}\Phi_{0 x} + \chi_{y}\Phi_{0 y}) + B &= 0 , \\
		B(\chi_{x}\Phi_{0 x} + \chi_{y}\Phi_{0 y}) - A\chi_{Z} &= 0 \label{lokbc1}. 
	\end{align}
\end{subequations}
Much as in the two-dimensional case, the final two equations give the non-trivial form of $\chi_{Z}$ on the free surface $Z=0$ as 
\begin{equation}
 \chi_{Z} = (\chi_{x}\Phi_{0 x} + \chi_{y}\Phi_{0 y})^{2} ,
\end{equation}
which we can then substitute into \eqref{eikonal3d1} to provide
\begin{equation} \label{3dnonlineikonal}
 	\chi_x^2 + \chi_y^2 + (\chi_x \Phi_{0x} + \chi_y \Phi_{0y})^4 = 0 .
\end{equation}
Analogous to the 2D case, we now argue that as a consequence of the chain rule, $\Phi_{0 x,y} = \phi_{0 x,y} + \eta_{x,y}\phi_{0 z}$. Since $\eta_{0} = 0$, to leading order we can replace the $\Phi_{0}$ terms with $\phi_{0}$, \emph{i.e.}
\begin{equation} \label{3dnonlineikonal_again}
 	\chi_{x}^{2}+\chi_{y}^{2}+(\chi_{x}\phi_{0 x} + \chi_{y}\phi_{0 y})^{4} = 0,
\end{equation}
which is the three-dimensional version of \eqref{2dnonlineikonal}.

Once more, we want to obtain ray equations via Charpit's method. Letting
\begin{equation}
\begin{gathered}
 	p=\chi_{x}, \quad q = \chi_{y} \\
	\Delta = p\phi_{0 x} + q\phi_{0 y}, \quad F = p^{2} + q^{2} +\Delta^{4} ,
\end{gathered}
\end{equation}
we find from \eqref{3dnonlineikonal_again},
\begin{gather}
\begin{aligned}
 	\dd{x}{\tau} & = 2p + 4\phi_{0 x}\Delta^{3} , &
	\dd{y}{\tau} & = 2q + 4\phi_{0 y}\Delta^{3} ,\\
	\dd{p}{\tau} & = -4\Delta^{3}(p\phi_{0 xx} + q\phi_{0 xy}), &
	\dd{q}{\tau} & = -4\Delta^{3}(p\phi_{0 xy} + q\phi_{0 yy}),\\
	\dd{\chi}{\tau} & = 2(p^{2} + q^{2} + 2\Delta^{4}) .
\end{aligned}
\end{gather}
We now note that as $F = 0$ from the eikonal equation,
\begin{equation}
 	\Delta^{4} = -p^{2} - q^{2} ,
\end{equation}
and hence
\begin{equation}
 	\dd{\chi}{\tau} = -2(p^{2}+q^{2}) .
\end{equation}
Ensuring we remember we are now only working in $x$ and $y$, we may concisely write this system using vector notation with 
\begin{equation}
 	\bm{x} = 
	\begin{pmatrix}
		x \\
		y
	\end{pmatrix} ,
	\qquad 
	\bm{p} = 
	\begin{pmatrix}
	 	p \\
		q
	\end{pmatrix} ,
\end{equation}
as 
\begin{subequations} \label{neatnlcheqns}
\begin{align}
 	\dd{\bm{x}}{\tau} & = 2\bm{p} + 4(\bm{p}\cdot\nabla\phi_{0})^{3}\nabla\phi_{0} , \\
	\dd{\bm{p}}{\tau} & = -4(\bm{p}\cdot\nabla\phi_{0})^{3}H\bm{p} , \\
	\dd{\chi}{\tau} & = -2(\bm{p}\cdot\bm{p}) ,
\end{align}
\end{subequations}
where $H$ is the Hessian matrix of second derivatives of $\phi_{0}$, \emph{i.e.}
\begin{equation}
 	(H)_{i j} = \pd{^{2}\phi_{0}}{x_{i}\partial x_{j}} .
\end{equation}

\section{Initial conditions for the source problem} \label{sec:initsource}

Derivation of initial conditions also follows much as before. 

Instead of a general geometry, we will illustrate the process for the case of three-dimensional uniform flow in the $x$-direction past a point source of (nondimensional) strength $\delta$. The source is placed at $(x,y,z) = (0,0,-h)$ in the fluid. A straightforward flux argument confirms that we must have $\delta < 0$ for a point source, while $\delta>0$ corresponds to a point sink. This flow is chosen to allow comparison of results (for nonlinear geometries) with those found in \cite{lustri2013steady} (for linear geometries), as this aspect of the comparison will be further discussed in Sec.~\ref{sec:3dlin}. 

In the low-Froude limit, the dominance of gravity causes the leading-order free surface to be flat, and hence we must insert an image source to the above point source. The leading-order velocity potential is therefore 
\begin{equation} \label{lovpsource}
 	\phi_{0} = x + \frac{\delta}{4\pi}\left\{\frac{1}{\sqrt{x^{2}+y^{2}+(z-h)^{2}}} +\frac{1}{\sqrt{x^{2}+y^{2}+(z+h)^{2}}} \right\}.
\end{equation}
The reader can refer to, \emph{e.g.}~\cite{milne1968theoretical} for further discussion about elementary potential flows similar to the above. So from $\phi_0$, we may observe that the singularities on the free-surface, taken as $z=0$, satisfy
\begin{equation}
 	x^{2}+y^{2}+h^{2} = 0 .
\end{equation}
These are precisely the singularities that cause a divergence of the late-order asymptotic terms. Thus we would like to choose the initial conditions for our ray equations as
\begin{equation} \label{pointsourceic}
 	x_{0}(s) = s \quad y_{0}(s) = \pm\im\sqrt{s^{2}+h^{2}} \quad \chi_{0}(s) = 0,
\end{equation}
with $s\in\mathbb{C}$ in general, denoting the variable that parameterises the initial condition. 

Using $F=0$ and the chain rule for $\de{\chi}/\de{s}$ we may find the equations for $p_{0}$ and $q_{0}$, as given by
\begin{align} \label{nlp0q0}
 	p_{0}^{2} + q_{0}^{2}+(p_{0}\phi_{0 x} + q_{0}\phi_{0 y})^{4} & = 0 ,\\
	p_{0} + \dd{y_{0}}{s}q_{0} & = 0 . \label{p0asq0eqn}
\end{align}
Assuming $y_{0}' \neq 0, \, \infty$ \emph{i.e.} $s\neq0 , \, \pm\im h$, we may use the second equation in the first to see
\begin{equation}
	p_{0}^{2}\left(1 + \frac{1}{y_{0}'^{2}}\right) +p_{0}^{4}\left(\phi_{0 x} - \frac{\phi_{0 y}}{y_{0}'}\right)^{4} = 0 ,
\end{equation}
and hence either $p_{0} = 0$ or 
\begin{equation} \label{nlpic}
 	p_{0} = \pm\im\frac{\left(1 + \frac{1}{y_{0}'^{2}}\right)^{1/2}}{\left(\phi_{0 x} - \frac{\phi_{0 y}}{y_{0}'}\right)^{2}} .
\end{equation}


Much as in Sec.~\ref{sec:2dic}, we have the problem that the ray equations \eqref{neatnlcheqns} rely on derivatives of $\phi_{0}$ -- again, for a numerical procedure we must begin slightly away from the singularity. As such, we let
\begin{equation} \label{nlsourceic}
 	x_{0, \rho, \nu} = s + \rho\e^{\im\nu}, \quad y_{0, \rho, \vartheta} = \pm\im\sqrt{s^{2}+h^{2}} + \rho\e^{\im\vartheta}, \quad \chi_{0} = \tilde{\delta} ,
\end{equation}
with $0<\rho\ll 1$, $0\le \nu, \vartheta < 2\pi$, and $\tilde{\delta} \to 0$ as $\rho \to 0$. As in Sec.~\ref{sec:2dic}, for practical implementation in Chap.~\ref{chap:num3d} we will take $\tilde{\delta} = 0$. We also need to approximate $y_{0}'$. Noticing that 
\begin{equation}
 	\dd{y_{0}}{s} = \pm\frac{\im s}{\sqrt{s^{2}+h^{2}}} = -\frac{x_{0}}{y_{0}} ,
\end{equation}
we choose to take
\begin{equation} \label{nly0'ic}
 	y_{0, \rho, \nu, \vartheta}' = -\frac{x_{0, \rho, \nu}}{y_{0, \rho, \vartheta}} ,
\end{equation}
and then under the assumption of integrability simply evaluate $p_{0}, \, q_{0}$ at these points. Were we to take $p_{0} = 0$, this would imply $q_{0} = 0$, and lead to a trivial solution for $\chi$. It follows that \eqref{nlpic} is the necessary condition for $p_{0}$, which we may then substitute into \eqref{p0asq0eqn} to find $q_{0}$.

\section{Linearised source problem \label{sec:3dlin}}

As previously mentioned, \cite{lustri2013steady} also consider the problem of flow past a point source. However, instead of transforming the system \eqref{chap1:nondim}, they proceed by assuming the source strength is sufficiently small, that is $|\delta|\ll\ep$, to allow linearisation about uniform flow -- \emph{i.e.} they use
\begin{equation}
 	\phi = x + \delta\bar{\phi} , \qquad \eta = \delta\bar{\eta} ,
\end{equation}
and only retain leading-order terms in $\delta$. For length considerations we omit the details, as after finding the linearised system it follows along much the same lines as for the nonlinear method described above. The resulting leading-order equation for $\chi$ on the linearised free surface, $z = 0$, is as follows (\emph{cf.} \eqref{3dnonlineikonal}): 
\begin{equation} \label{lfseikonal}
 	\chi_{x}^{4} + \chi_{x}^{2} + \chi_{y}^{2} = 0 .
\end{equation}
Seeking ray solutions, we again apply Charpit's method and hence find the corresponding equations for the problem as
\begin{gather} \label{linsourcecharp}
\begin{aligned}
 	\dd{x}{\tau} & = 4p^{3} + 2p , &
 	 \dd{y}{\tau} & = 2q,\\
 	\dd{p}{\tau} & = 0 , &
	\dd{q}{\tau} & = 0 ,\\
 	\dd{\chi}{\tau} & = 2p^{4} .
\end{aligned}
\end{gather}
Important to note is that as \eqref{lfseikonal}, $p^{4} + p^{2} + q^{2} = 0$, has no explicit $x$- or $y$-dependence, we see that $p$ and $q$ are  constant along the rays. As the other equations solely depend on these terms, it follows that our rays are now straight. 

The leading-order velocity potential for the problem remains \eqref{lovpsource} -- as such, the singularity location is identical, but the issue of starting at the singularity itself has now been removed. Hence we may directly apply the initial conditions
\begin{equation} \label{linsourceic1}
 	x_{0}(s) = s, \quad y_{0}(s) = \pm\im\sqrt{s^{2}+h^{2}} , \quad \chi_{0}(s) = 0 ,
\end{equation}
along with $p_{0}(s)$ and $q_{0}(s)$ satisfying
\begin{equation} \label{linsourcepqic}
 	p_{0}^{4} + p_{0}^{2} + q_{0}^{2} = 0, \quad p_{0} + q_{0}\dd{y_{0}}{s} = 0.
\end{equation}
Once more, the second equation here provides that for $y_{0}' \neq 0, \, \infty$,
\begin{equation} \label{linsourceqaspic}
 	q_{0} = -\frac{p_{0}}{y_{0}'} ,
\end{equation}
and hence either $p_{0} = 0$ or
\begin{equation} \label{linsourcepic}
 	p_{0} = \pm\im\left(1 + \frac{1}{y_{0}'^{2}}\right)^{1/2} .
\end{equation}
However, note that if we choose $p_{0} = 0$, $\chi \equiv 0$, and hence for non-trivial solutions we must choose \eqref{linsourcepic}. The initial condition for $q_{0}$ then again follows from \eqref{linsourceqaspic}.

In fact, this problem is simple enough to solve analytically: As $p$ and $q$ are constant in $\tau$, we may use the initial conditions to find
\begin{equation} \label{linsourcepqsol}
 	p = \pm\frac{h}{s}, \quad q = \pm \frac{\im h(s^{2}+h^{2})^{1/2}}{s^{2}} ,
\end{equation}
where the sign of $q$ follows from those chosen for $y_{0}$ and $p$. Substituting these into \eqref{linsourcecharp}, we can find $x, \, y$ and $\chi$ in terms of $s$ and $\tau$, then invert to find $\chi$ in terms of $x$ and $y$. This is how \cite{lustri2013steady} proceed, finding eight possible singulant expressions. Four of these expressions satisfy the conditions \eqref{dinglecdn} on the line $x = 0$ irrespective of $h$, and hence this corresponds to a Stokes line -- we will return to this in Chap.~\ref{chap:num3d}. 


\section{Discussion}
In this chapter, we have introduced critical elements for the rest of our undertaking. Following work in \cite{chapman2006exponential}, we have presented the idea of a function, $\chi$, for which we may apply the conditions \eqref{dinglecdn} to find Stokes lines, across which exponentially small terms switch-on or -off. Extending a linear method developed in \cite{lustri2013steady}, we have then derived the equations necessary to determine this function for fully nonlinear steady potential flows in 2D,  \eqref{2dnonlineikonal}, and 3D, \eqref{3dnonlineikonal_again} -- at least on the free surface. In order to make these equations amenable to a numerical procedure, we applied Charpit's method to produce complex ray-theoretic equations \eqref{2dcharpiteqns} and \eqref{neatnlcheqns} respectively, and discussed the necessary initial conditions. In Sec.~\ref{sec:3dlin} we also display the equations for the linearised problem, showing clearly the substantive differences -- the rays are straight rather than curved; the ability to start a corresponding numerical procedure from the singularity; the ability to solve the system analytically.


\chapter{Stokes lines for two-dimensional flow via direct integration \label{chap:raytheoryint}}

In the previous chapter we derived the 2D ray equations, \eqref{2dcharpiteqns}, to allow approximation of the singulant $\chi$. As explained, this allows us to determine the Stokes lines in the fluid by using the \cite{dingle1973asymptotic} condition, \eqref{dinglecdn}. However, we have the means to proceed more straightforwardly. Returning to \eqref{2dnonlineikonal}, rather than forming ray equations we may instead simply observe that as $\chi$ is non-constant, 
\begin{equation} \label{chix}
	\chi_x = \pm\frac{\im}{\phi_{0x}^2} .
\end{equation}
This allows us to integrate directly to find $\chi$ on the free surface, $z\approx 0$, so that
\begin{equation} \label{chiintz}
	\chi_{\pm}(x, 0) = \pm \im \int_{x_0}^x \frac{1}{\phi^2_{0x}(s, 0)} \, \de{s} ,
\end{equation}
where if possible $x_{0}$ is chosen such that $\chi(x_{0},0)=0$, \emph{i.e.} at one of the singularities of the leading-order solution. In fact due to the uniqueness of analytic continuation, if we allow $x\in\mathbb{C}$ we may identify $\chi$ as defined by \eqref{chiintz} with $\chi$ in the physical fluid domain. This holds so long as we remain on the same Riemann sheet as the physical fluid.

In 3D, we have no choice 
but to use our systems of ray equations, \eqref{neatnlcheqns} or \eqref{linsourcecharp}. This two-dimensional situation, where we may use the integral \eqref{chiintz} to calculate $\chi$, allows the opportunity to obtain Stokes lines for comparison with those from the numerical methods of Chap.~\ref{chap:num2d}. It will also allow us to derive the local behaviour of $\chi$ for a chosen singularity, to be used in the same chapter. 

\section{Flow over a step \label{sec:fos}}

\subsection{Finding the Stokes lines}

An oft-considered problem is that of flow over a rectangular step, as in \emph{e.g.} \cite{king1987free,chapman2006exponential,trinh2016topological}. 

%
We use the boundary-integral method introduced in Sec.~\ref{sec:boundint}. After nondimensionalisation, the necessary conformal map is given by 
\begin{equation} \label{conformalmapper}
 	\zeta = \xi + \im\eta = \e^{-w} .
\end{equation}
For a visualisation of this problem in the $z$- and $\zeta$-planes, see \cite{king1987free},~Fig.~1. Omitting derivation for the sake of brevity (\emph{cf.} \cite{chapman2006exponential}, Sec.~{5}), \eqref{chap1:boundintsys} provides that on the complexified free surface $\xi \in \mathbb{C}$, $q$ and $\theta$ satisfy
\begin{subequations} \label{compqeqns}
	\begin{align}
 		\log q(\xi) \mp \im\theta(\xi) &=  \log \left(\frac{\xi + b}{\xi + c}\right)^{1/2} \hspace{-0.04in} -\frac{1}{\pi} \int_{0}^\infty \frac{\theta(\xi')}{\xi' - \xi} \, \de{\xi'} , \label{compqint} \\
		\epsilon\xi q^{2}\dd{q}{\xi} &= \sin\theta, \label{compqde}
	\end{align}
\end{subequations}
where recall that the $\mp$ sign corresponds to $\xi$ in the UHP/LHP. Here, 
$-b, \, -c$ are respectively the images of the lower- and upper-corners of the step under the map \eqref{conformalmapper}.
Now taking the low-Froude limit, the leading-order expressions are hence
\begin{equation} \label{stepqthetalo}
 	q_{0} = \left(\frac{\xi +b}{\xi + c}\right)^{1/2} , \qquad \theta_{0} = 0 .
\end{equation}

From the equation for $q_{0}$, we see that the important singularity where $\chi = 0$ corresponds to $\xi = -c$. 

We now wish to use \eqref{chiintz} to find $\chi$ on the complexified free surface. Recall that so long as we remain on the Riemann sheet of the physical fluid we may identify $\xi \in \mathbb{C}$ with the original $\zeta$. 
Nonetheless, to prevent confusion we will choose to label all two-dimensional graphs in terms of the relevant complexified free-surface variable. 

\begin{figure}[htb] \centering
	\includegraphics[width=0.7\textwidth]{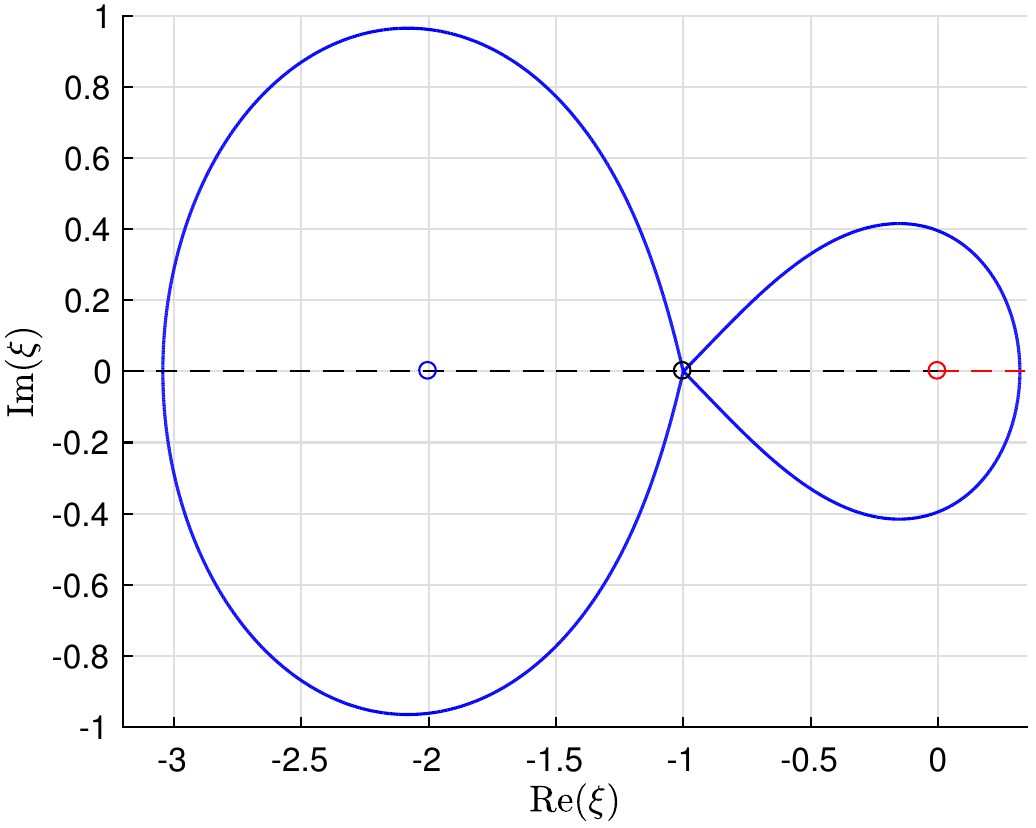}
	\caption{Approximate Stokes lines for the flow in the complex $\xi$-plane (or equivalently the $\zeta$-plane) using $\chi_{\pm}$ found from equation \eqref{chiintxi} with $c = 1, \, b = 2$. We see that the Stokes lines for the two singulants overlay each other. Provided we are on the correct Riemann sheet, the black dashed line corresponds to the real bottom, the red dashed line to the real free surface. Also marked with circles are the singularities of $\chi$, $\xi = -b$ (blue) and $\xi = 0$ (red). \label{stokeslinesfoszp}}
\end{figure}

Now, as $\phi_{0x}$ is simply the leading-order horizontal speed and we have found $\theta_{0} = 0$, we may approximate this as $q_{0}$. To change the integration variable from $x$ to $\xi$, we use that by \eqref{qtheta1}
\begin{equation} \label{fosdxdw}
 	\dd{x}{w} \approx \frac{1}{q_{0}} ,
\end{equation}
and by \eqref{conformalmapper}, $w'(\xi) = -1/\xi$. Hence, by the chain rule we obtain
\begin{equation} \label{chiintxi}
 	\chi_{\pm}(\xi) = \mp\im \int_{-c}^{\xi} \frac{1}{\xi' q_{0}^{3}} \de{\xi'}.
\end{equation}
The resulting Stokes lines (as defined by the conditions \eqref{dinglecdn}), are shown in the $\xi$-plane for $c = 1, \, b = 2$ in Fig.~\ref{stokeslinesfoszp}.  

Important to observe from \eqref{chiintxi} is that $\chi$ contains singularities at $\xi = -b$, $\xi = 0$. Should the contour of integration progress around one of these, it will change Riemann sheet and thus we will lose correspondence with the real fluid domain. 
We defer to \cite{trinh2013new2} for a more rigorous treatment of this problem, and \cite{trinh2016topological},~Fig.~5 for a visualisation of the Riemann surface structure. We also note that should it be desired, we may integrate \eqref{fosdxdw} to find $x$ in terms of $w$. The constant of integration is determined by ensuring that the corner of the step corresponds to the leading-order singularity, $w = -\log(c) - \im\pi$. Having done so, it is  possible (at least numerically) to project the Stokes lines found in Fig.~\ref{stokeslinesfoszp} into the real fluid domain -- doing so results in  \cite{chapman2006exponential},~Fig.~3, but for length considerations we omit the process here.

\subsection{Local analysis \label{foszplocan}}

For the numerical methods described in Chaps.~\ref{chap:num2d}~and~\ref{chap:num3d}, we will see that our starting location is highly important. As such, here we seek the local behaviour of $\chi$ near the singularity $\xi = -c$. As $\xi \to -c$, from \eqref{stepqthetalo} we have that
\begin{equation}
 	q_{0} \sim (\xi + c)^{-1/2} ,
\end{equation}
and so substituting this into \eqref{chiintxi} and using $\xi' \approx -c$ we find
\begin{equation}
 	\chi_{\pm} \sim \pm\im\frac{2}{5c}(\xi + c)^{5/2} .
\end{equation}
Letting $\xi + c = \rho\e^{\im\theta}$, we see that to meet the condition $\Im(\chi) = 0$ we require 
\begin{equation}
 	\frac{5\theta + \pi}{2} = n\pi \quad n\in\mathbb{Z} ,
\end{equation}
and hence $\theta = (2n-1)\pi/5$. Note that for these to be Stokes lines, they must also satisfy the additional condition $\Re(\chi) \ge 0$, but every $n$ will provide a Stokes lines for one of $\chi_{\pm}$. 

\section{Flow over a nonlinear geometry \label{sec:nls}}

\begin{figure}[htb] \centering
	\includegraphics[width=0.8\textwidth]{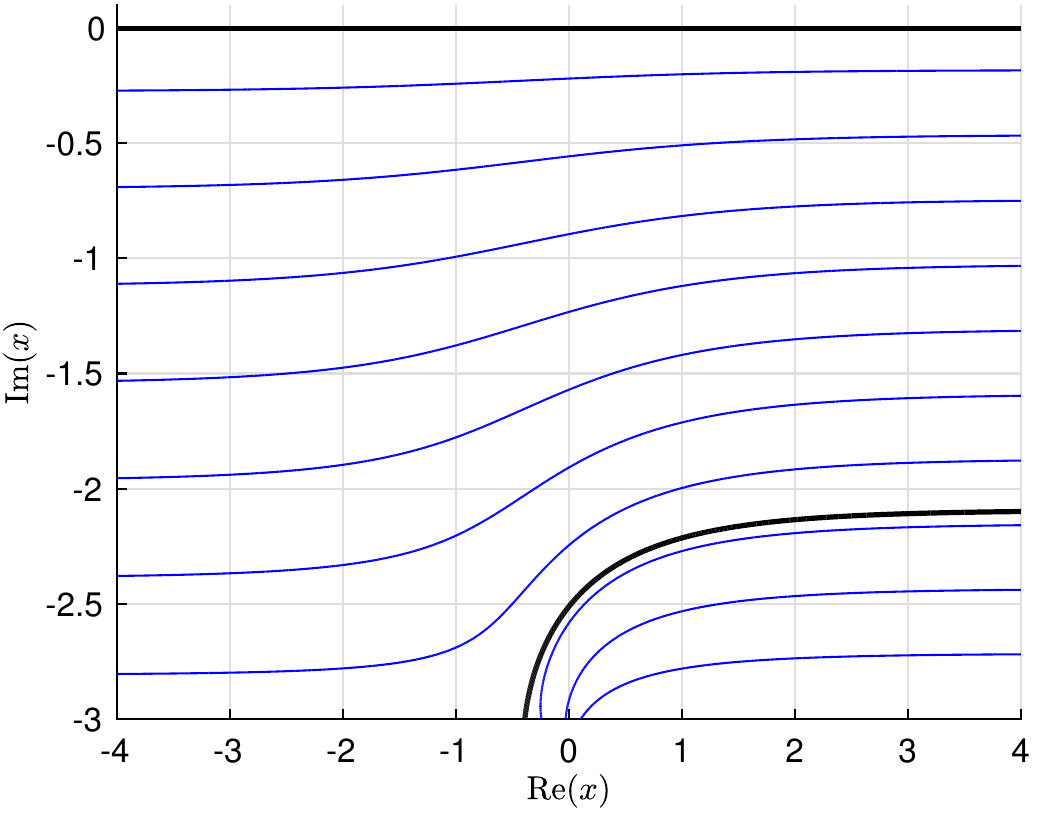}
	\caption{Creating a nonlinear geometry by taking the streamline $\Im(\phi_{0}) = -\pi$ in \eqref{nonlinphi0} as a rigid wall (the lower thick black curve). Blue lines are other streamlines of the flow -- those beneath $\Im(\phi_{0}) = -\pi$ would not be physically relevant.  \label{nonlingeom}}
\end{figure}

We now assume the step to be vanishingly small, setting $c = 1$ and $b = 1 + \delta$, where $\delta \ll 1$ (\emph{cf.} \cite{trinh2013new}). Linearising about the uniform flow we see that to leading order in $\delta$, $q = 1 + \delta\tilde{q}$, $\theta = \delta\tilde{\theta}$. Furthermore, taking the low-Froude limit and expanding $\tilde{q}, \, \tilde{\theta}$ in powers of $\ep$, from \eqref{qtheta1} we have 
\begin{equation} \label{fossdwdz}
	\dd{w}{z} = 1 + \delta (\tilde{q}_0 - \im\tilde{\theta}_0) + \Oh(\delta^2, \delta\epsilon) .
\end{equation}
We may perform the same expansions in \eqref{compqeqns}, and observe 
\begin{equation}
 	\tilde{q}_{0} = \frac{1}{2(\xi + 1)} , \qquad \tilde{\theta}_{0} = 0 .
\end{equation}
To transform into the $z$-plane we use that $w = z + \Oh(\delta)$, and thus (identifying $\xi \in \mathbb{C}$ with the original $\zeta$) $\xi = \e^{-z + \Oh(\delta)}$. Substituting this into \eqref{fossdwdz} and integrating then provides that 
\begin{equation}
	w(z) = z + \frac{\delta}{2} \log [2(1 + \e^{z})] + \Oh(\delta^2, \delta\epsilon) .
\end{equation}
Taking the real part, we may find our leading-order velocity potential, $\phi_{0}$. To find Stokes lines we solely require $\phi_{0}$ on the complexified free surface, $x \in \mathbb{C}, \, y = 0$ -- here 
\begin{equation} \label{nonlinphi0}
 	\phi_{0}(x, 0) = x + \frac{\delta}{2}\log [2(1 + \e^{x})] ,
\end{equation}
and so singularities lie at $x = \im\pi(2n + 1), \, n \in \mathbb{Z}$. The correspondence of this expression for $x \in \mathbb{C}$ with $w(z)$ is clear. While we have obtained this equation by assuming $\delta \ll 1$, we will now set $\delta = 1$ -- this new potential may be thought of as a superposition of an infinite number of point sources, placed at each singularity location. Additionally, we could identify one of the streamlines of the flow with a rigid boundary, and so instead consider this to be flow past a nonlinear geometry. This is pictured in Fig.~\ref{nonlingeom} for the streamline $\Im(\phi_{0}) = \psi = -\pi$. 

\subsection{Finding the Stokes lines \label{sec:nlsstokes}}

\begin{figure}[htb] \centering
	\includegraphics[width=0.6\textwidth]{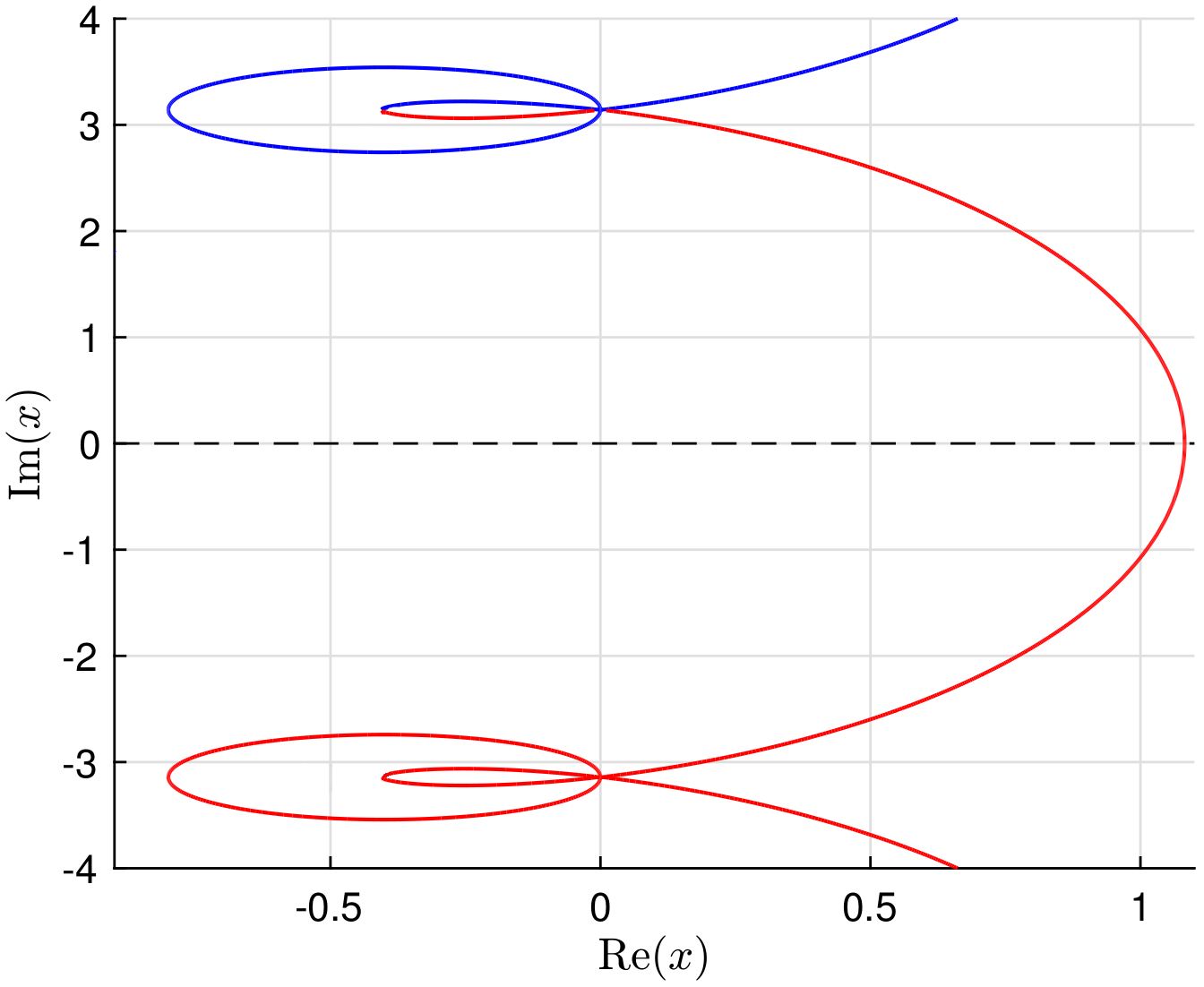}
	\caption{Approximate Stokes lines from the singularity at $x=\pi\im$ for the flow with $\phi_{0}$ given by \eqref{nonlinphi0} with $\delta = 1$.  For $\chi_{\pm}$ given by \eqref{chiintz} -- shown blue for $\chi_{+}$, red for $\chi_{-}$ -- we see there is a Stokes line hitting the real free surface $\Im(x) = 0$ (shown dashed black), corresponding to the negative sign. \label{stokeslinesnonlin1}}
\end{figure}

As suggested in Chap.~\ref{chap:expascrt}, we expect the dominant contributions to come from those singularities closest to the real free surface \citep{chapman2005exponential}, \emph{i.e.} $x = \pm\im\pi$. Considering the singularity $x = \im\pi$, substituting 
\begin{equation} \label{nonlinphi0x}
 	\phi_{0 x} = 1 + \frac{\e^{x}}{2(1 + \e^{x})}
\end{equation}
into \eqref{chiintz} and using the conditions \eqref{dinglecdn} results in the Stokes lines shown in Fig.~\ref{stokeslinesnonlin1}. We see that taking the negative sign in \eqref{chiintz} results in free surface intersection. An analogous picture applies for the conjugate singularity $x = -\im\pi$, now with the positive sign providing intersection. By the Schwarz reflection principle (see \emph{e.g.} \cite{krantz2012handbook}) we expect this to correspond to a conjugate term switching-on, thus resulting in real waves (as argued in several papers for free-surface waves, \emph{e.g.} \cite{trinh2013new}).

\subsection{Local analysis \label{nlslocan}}

We now investigate the local behaviour near the singularity $x = \im\pi$, for later use in Chap.~\ref{chap:num2d}. Substituting in \eqref{nonlinphi0x} to \eqref{chix}, we have 
\begin{equation}
 	\chi_{x} = \pm\frac{\im}{\phi_{0 x}^{2}} = \pm\frac{4\im(1+\e^{x})^{2}}{(2+3\e^{x})^{2}} .
\end{equation}
%
Letting $X = x - \im\pi$, near the singularity $X \to 0$, $\phi_{0 x} \sim 1/(2X)$, and hence
\begin{equation} \label{locanchix}
 	\chi_{x} \sim \pm 4\im X^{2} ,
\end{equation}
which we may integrate and apply $\chi(\im\pi) = 0$ to find that
\begin{equation} \label{locanchi}
 	\chi \sim \pm \frac{4\im}{3} X^{3} .
\end{equation}
%
We note that an analogous local analysis can be applied to Charpit's equations \eqref{2dcharpiteqns}, and provides the same local behaviour. The corresponding choice of sign is simply that taken for $p_{0}$ in \eqref{2dp0ic}. 

If we now take $X = \rho\e^{\im\theta}$, we see that for $\Im(\chi) = 0$ we require
\begin{equation}
 	\frac{6\theta + \pi}{2} = n\pi , \quad n\in\mathbb{Z} ,
\end{equation}
and hence $\theta = (2n - 1)\pi/6$. Again, these equal-phase lines must satisfy the additional condition $\Re(\chi) \ge 0$ to be Stokes lines, but each $n$ will correspond to a Stokes line for one of the choices of sign above.

\section{Discussion}

In this chapter we have used the ability to directly integrate \eqref{chix} to find Stokes lines in 2D. We will return to these graphs for comparison with those produced by our numerical method in the subsequent chapter. We also performed local analyses to find the angles of the equal-phase lines at the singularity for both problems considered, again for later use. As noted, the use of a complex boundary-integral formulation such as \eqref{compqeqns} is restricted to 2D by the reliance on complex variable methods (such as conformal mapping). It is for this reason that it is not of particular importance to this dissertation, as we desire to find Stokes surfaces for three-dimensional flows. The potential considered in Sec.~\ref{sec:nls} provides an example of how we may use point sources to model nonlinear bottom geometries (\emph{cf.} Fig.~\ref{nonlingeom}), and stimulates the possibility of using analogous potentials for 3D obstructions. We will study the case of a single nonlinear three-dimensional point source in Chap.~\ref{chap:num3d}.

%
\chapter{Numerical Stokes lines in 2D \label{chap:num2d}}

In Chap.~\ref{chap:raytheoryint}, we used that in 2D it is possible to use the integral \eqref{chiintz} to find $\chi$, and then after apply the conditions \eqref{dinglecdn} to find Stokes lines. However, it will not always be possible to solve for $\chi$ analytically, and for our nonlinear 3D ray equations, \eqref{neatnlcheqns}, the dependence on the gradient of the leading-order velocity potential will almost certainly prevent us from doing so. As a result, we must develop a numerical procedure that can directly find the Stokes lines (or in 3D, surfaces). In the following two chapters we present four different methods to do so (summarised in pseudocode in App.~\ref{app:pseudo}), each deduced from the base logic of Sec.~\ref{sec:numbasis} but applicable to different situations. 

As suggested in Chap.~\ref{chap:expascrt}, the key to numerically finding the Stokes lines/surfaces is in satisfying the equal-phase condition -- once this is accomplished the additional condition of $\Re(\chi)\ge 0$ is easily checked. 

\section{Basis of numerical method for Stokes lines \label{sec:numbasis}}

Following \cite{autosteep}, we begin by considering $\chi$ in terms of the characteristic variable, $\tau$, and suppose that the equal-phase lines are given by
\begin{equation}
 	V(\tau) = \Im(\chi) = \text{constant} ,
\end{equation}
for some differentiable function $V$. Given an initial point $s\in\mathbb{C}$, we want to choose $\tau=\tau_{1}(t)+\im\tau_{2}(t)$ such that the ray solution, $\chi(\tau(t); \, s)$, has $\Im(\chi(t))$ constant. As $\nabla V$ in the $(\tau_{1},\tau_{2})$-plane is perpendicular to the level sets of V, along the equal-phase lines the position vector $\bm{r}(t) = (\tau_{1}(t), \tau_{2}(t))$ must satisfy
\begin{equation} \label{tautperpv}
 	\dd{\bm{r}}{t}\propto(\nabla V)^{\perp} = \left(-\frac{\partial V}{\partial \tau_{2}},\frac{\partial V}{\partial\tau_{1}}\right) .
\end{equation}
Further, under the assumption that $\chi$ is holomorphic away from critical points of the flow we have 
\begin{equation}
 	\dd{\chi}{\tau} = \pd{V}{\tau_{2}} + \im \pd{V}{\tau_{1}} ,
\end{equation}
and hence by comparing this with \eqref{tautperpv} we have the necessary relation for $\tau$. As we are free to scale $t$, we additionally choose to normalise so that 
\begin{equation} \label{tautaschitau}
 	\dd{\tau}{t} = -\overline{\dd{\chi}{\tau}}\biggr/\left|\dd{\chi}{\tau}\right| ,
\end{equation}
where the bar denotes complex conjugation, and the initial condition to take is simply $\tau(0) = \tau_{0}$ (determined by the problem under consideration). Using the chain rule, we may hence numerically solve our original system by integrating with respect to $t$. Notice that in applying the level set scheme of \eqref{tautperpv}, we could have negated the RHS. Our choice of sign has the effect that in the subsequent methods $\Re(\chi)\ge 0$ for decreasing values of $t$.

Recalling from Chap.~\ref{chap:raytheoryint} that in 2D we may identify the complexified free surface with the real plane, it follows that we may numerically solve for the Stokes lines within the fluid. However, in 3D only at the intersection of the Stokes surface with the real free surface do we know the correspondence between our analytically continued variables and physical space. As such, for use in our numerical procedures we define the measure
\begin{equation} \label{fsmeasure}
 	\mu(\bm{x}) = \sum_{i} |\Im(x_{i}) | ,
\end{equation}
so that $\mu(\bm{x}) = 0$ corresponds to intersection. In practice, due to numerical inaccuracy this condition may have to be relaxed slightly, at least in 3D. 

A final condition that must be satisfied is that our leading-order velocity potential is real when $\mu(\bm{x}) = 0$ -- this is to ensure we lie on the correct Riemann sheet, which corresponds to that of the physical fluid. Consequently, we have the additional equation to evolve with the rest of the system, 
\begin{equation} \label{phi0teqn}
 	\dd{\phi_{0}}{t} = \nabla\phi_{0}\cdot\dd{\bm{x}}{\tau}\dd{\tau}{t} ,
\end{equation}
with the initial condition $\phi_{0}(\bm{x}_{0})$. From this we obtain the $\phi_{0}$ value on the ray, and simply require $\Im(\phi_{0}) = 0$ when $\mu(\bm{x}) = 0$. Again, in practice this condition may have to be relaxed. 

There is an additional problem present, which is that it may be difficult to ensure that the terms with which we are evolving our system are continuous throughout the integration (\emph{e.g.} due to the path crossing a branch cut). To compensate for this, as in \cite{autosteep} we should convert the terms to a combination of power or logarithmic functions (more general functions are treated similarly), and change the problem to a larger system of ODEs. A concrete example of what this means, and further explanation as to why is given in Sec.~\ref{numchap:step}.

\section{Two-dimensional method via integral equation \label{sec:2dintm}}

As previously stated, the idea used in Chap.~\ref{chap:raytheoryint} -- first integrating to find $\chi$, and then seeking Stokes lines where the conditions \eqref{dinglecdn} hold -- may not always be possible (or at least practical). In this section, we apply the logic of Sec.~\ref{sec:numbasis} to form a numerical method which instead, starting from a given integrand, directly solves for the Stokes lines. We will explain the method for \eqref{chiintz}, where the integration variable is $x$, but as used in Sec.~\ref{numchap:step}, it is applicable more broadly. Recalling \eqref{chiintz}, on the complexified free surface we have 
\begin{equation} \label{numdchidx}
 	\dd{\chi}{x} = \pm\frac{\im}{\phi_{0 x}^{2}(x,0)} , 
\end{equation}
and hence as $x$ is now the dependent variable, by \eqref{tautaschitau} we must take
\begin{equation} \label{numintxt}
 	\dd{x}{t} = \pm\left(\frac{1}{\overline{\phi_{0 x}^{2}}(x(t),0)}\right)|\phi_{0 x}^{2}(x(t),0)| ,
\end{equation}
along with 
\begin{equation} \label{numintcpt}
 	\dd{\chi}{t} = \dd{\chi}{x}\dd{x}{t}, \qquad \dd{\phi_{0}}{t} = \phi_{0 x}\dd{x}{t} .
\end{equation}
%
Much as for the ray methods in Chap.~\ref{chap:expascrt}, the dependence on $\phi_{0 x}$ means a numerical solver will have difficulty starting exactly on any chosen singularity, $x_{0}$ say. As such, we will use the initial conditions shown in \eqref{2dcharpic}, along with $\phi_{0} = \phi_{0}(x_{0,\rho,\nu})$. 
When solving this system in practice, bounds on $|\bm{x}|$ and $|\chi'(x)|$ may be necessary to prevent the numerical solver leaving the domain of interest or entering critical points where $|\chi'(x)| \to \infty$. 

The method to find the Stokes lines intersecting the real free surface is then as follows:
\begin{enumerate}
	\item Determine singularities of the leading-order velocity potential on the complexified free surface. 
	\item Choose a singularity, recalling from Chap.~\ref{chap:expascrt} that the most important is typically that which is closest to the real axis. 
	\item If necessary, perform the conversion of terms discussed to ensure continuity, then set initial conditions for \eqref{numintxt} and \eqref{numintcpt} as described above.  In doing so, a choice of $\rho$ and $\nu$ has been made, along with the choice of sign in \eqref{numdchidx}. As long as $\rho$ is taken sufficiently small, we would not expect it to have an impact on the convergence of the method to the Stokes line. However, we would predict that given a choice of Stokes lines, our procedure will converge to that closest to its initial condition. As changing $\nu$ can change which line this is, we will vary over a discrete set of values, 
	\begin{equation}
		\Theta_{n} = \{\nu_{0}, \nu_{1}, \dots, \nu_{2n - 1}\} , 
	\end{equation}
	where we take $\nu_{j} = j\pi/n$ for some $n \in \mathbb{N}$. 
	\item For the chosen initial conditions, put the system into a numerical integration program (\emph{e.g.} ode113 in \textsc{Matlab}), and solve for decreasing $t$ (so as to satisfy the additional condition $\Re(\chi)\ge 0$) up to a given maximum value. Also impose the conditions that the solver stop when either $\Im(x) = 0$ or chosen bounds on $|x|$ and $|\chi'(x)|$ are exceeded.
	\item For each ray that has terminated with $\Im(x) = 0$, ascertain if ${\Im(\phi_{0}) = 0}$ at the end point (to within a chosen tolerance).
	\item If these additional conditions are met, then plot the ray.
\end{enumerate}
Algorithm~\ref{alg:2dint} in App.~\ref{app:pseudo} shows a pseudocode for this method.

\section{Flow over a step \label{numchap:step}}

In Sec.~\ref{sec:fos}, we succeeded in finding Stokes lines in the complex $\xi$-plane for the problem of flow over a rectangular step (shown in Fig.~\ref{stokeslinesfoszp}). In this section we will use the method described in Sec.~\ref{sec:2dintm} to find these lines directly, without first solving for $\chi$ analytically. Recalling our integral equation for $\chi(\xi)$ given in \eqref{chiintxi}, and our leading-order expression for $q$ in \eqref{stepqthetalo},
\begin{equation} \label{fosqnum}
 	q_{0} = \left(\frac{\xi + b}{\xi + c}\right)^{1/2} , 
\end{equation}
 we have that
\begin{equation} \label{foschixipm}
 	\dd{\chi_{\pm}}{\xi} = \mp\im\frac{1}{\xi}\left(\frac{\xi + c}{\xi + b}\right)^{3/2} .
\end{equation}
Before proceeding further, we should hesitate as clearly there are branch cuts present. As suggested in Sec.~\ref{sec:numbasis}, these cuts mean that our numerical solver may not provide us with a continuous solution -- we should convert the problem to a combination of (in this case) power functions. For the current situation, we do so by introducing the auxiliary functions
\begin{equation} \label{fosauxfunc}
 	f_{1}(\xi) = (\xi + c)^{1/2}, \quad f_{2} = (\xi + b)^{-1/2} ,
\end{equation}
so that 
\begin{equation} \label{dchidxiaux}
	f_{1}'(\xi) = \frac{1}{2f_{1}} , \quad 
	f_{2}'(\xi) = -\frac{1}{2}f_{2}^{3} , \quad
 	\chi'(\xi) = \mp\im\frac{f_{1}^{3}f_{2}^{3}}{\xi} .
\end{equation}
The key idea of this process is that after setting initial conditions $f_{1}(\xi_{0}), \, f_{2}(\xi_{0})$ specifying the initial Riemann sheet, the right-hand sides of these equations are well-defined and single-valued everywhere apart from singularities. As a result, our solution will be single-valued also -- the branch cuts are effectively \emph{automatically} placed out of the way of the path. 

\begin{figure}[htb] \centering
	\includegraphics[width=0.73\textwidth]{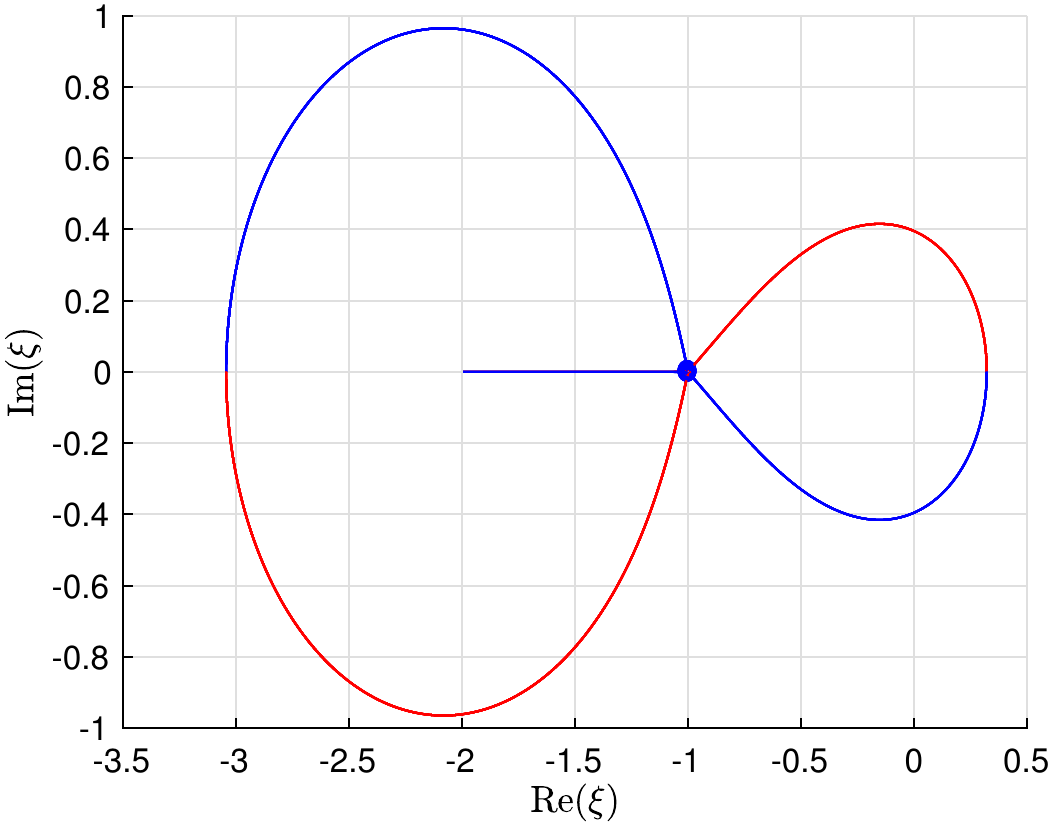}
	\caption{Numerical approximations to the Stokes lines in the complex $\xi$-plane for flow over a step. Starting points are given by $\xi = -1+\rho\e^{\im\nu_{j}}$ for $\rho=0.01$, $\nu_{j} \in \Theta_{16}$. With $\chi_{\pm}'(\xi)$ given by \eqref{foschixipm}, blue corresponds to $\chi_{+}$, red to $\chi_{-}$. Note that in comparison to Fig.~\ref{stokeslinesfoszp}, the two distinct colours stop abruptly because we limit the solution to a single Riemann sheet. \label{zetafosnum}}	
\end{figure}

\begin{figure}[htb] \centering
	\includegraphics[width=0.73\textwidth]{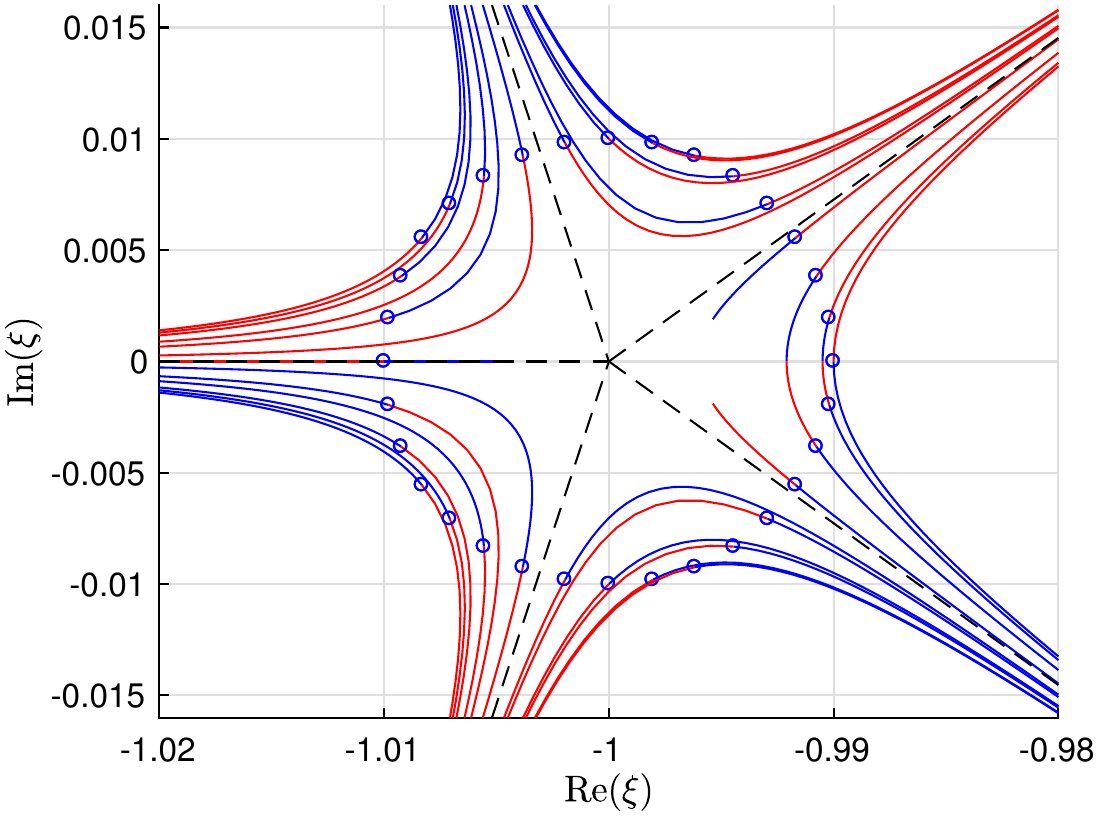}
	\caption{A closer look at the starting points in Fig.~\ref{zetafosnum}, shown as blue circles. Shown dashed black are the equal-phase lines $\theta = (2n - 1)\pi/5$, $n \in \mathbb{Z}$, found from local analysis in Sec.~\ref{foszplocan}. \label{zetafosnumzoomed}}	
\end{figure}

Moving forward, we evolve using
\begin{equation}
 	\begin{aligned}
	 	\dd{\xi}{t} & = -\overline{\dd{\chi}{\xi}}, &&& \dd{\chi}{t} &= \dd{\xi}{t}\dd{\chi}{\xi} , \\[4pt]
		\dd{q}{t} & = \dd{\xi}{t}\dd{q}{\xi}, &&& \dd{\mathbf{f}}{t} &= \dd{\xi}{t}\dd{\mathbf{f}}{\xi} .
	\end{aligned}
\end{equation}
Here $\mathbf{f} = (f_{1}, \, f_{2})$, $\chi'(\xi)$, and $\mathbf{f}'(\xi)$ are as in \eqref{dchidxiaux}, and we use the derivative of \eqref{fosqnum} in terms of the auxiliary functions for $q'(\xi)$. Note the slight deviation from the given method in Sec.~\ref{sec:2dintm}. There, we verified that $\phi_{0}$ was real on the free surface in order to ensure physicality. In this application, we instead solve for $q$ along the contour and then require $q > 0$  on the free surface -- here given by the positive $\Re(\xi)$-axis. The resulting numerical Stokes lines for $c = 1$, $b = 2$, are shown in Fig.~\ref{zetafosnum}, including those which do not hit the free surface. Comparing this with Fig.~\ref{stokeslinesfoszp}, we see that the method has performed very well. The region immediately around the singularity $\xi = -1$ is shown in Fig.~\ref{zetafosnumzoomed}, along with the equal-phase lines we found by local analysis in Sec.~\ref{foszplocan}. 

Several important details should be noted. Firstly, while the additional line from $\xi = -1$ to $-2$ is a Stokes lines, it lies along a branch cut, and is not involved in a physically relevant switching. From Fig.~\ref{zetafosnumzoomed}, we can observe that the line the method converges to is determined by the sector from which it begins -- bounded on either side by the equal-phase lines where $\Re(\chi)\le 0$. This is precisely the behaviour predicted, but we can additionally see that there is slower convergence near the boundaries, and in fact no convergence whatsoever if we begin on the incorrect phase line itself. So long as we run the method for a sufficiently large number of initial conditions, this will not present any difficulties. 
%
%
\vspace{0.4in}

\section{Two-dimensional method via characteristic equations \label{sec:2dcharpm}}
We also desire a method to find the Stokes lines directly from the two-dimensional ray equations given in \eqref{2dcharpiteqns}. This follows the same steps as in Sec.~\ref{sec:2dintm}, but with $\tau$ as the dependent variable. Our differential system in $t$ is obtained by applying the chain rule, \emph{i.e.} using \eqref{tautaschitau} we have 
\begin{equation} \label{2ddtaudt}
 	\dd{\tau}{t} = -\overline{p^{2}}/|p^{2}| ,
\end{equation}
and then
\begin{equation} \label{numchap:2dcharpitseqns}
	\dd{x}{t} = \dd{\tau}{t}\dd{x}{\tau}, \qquad
	\dd{p}{t} = \dd{\tau}{t}\dd{p}{\tau} , \qquad
 	\dd{\chi}{t} = \dd{\tau}{t}\dd{\chi}{\tau} ,
\end{equation}
with appropriate initial conditions as discussed in Sec.~\ref{sec:2dic}, and $\tau_{0} = 0$. Much as for the integral method, in practice it may be necessary to place bounds on $|x|$ and $|p|$. Additionally, we must include the corresponding equation for $\phi_{0}$, if necessary along with those for auxiliary functions stemming from conversion of terms. The procedure to find the Stokes lines is then analogous to that in Sec.~\ref{sec:2dintm}, so we will not explicitly state it here -- Algorithm~\ref{alg:nl2dcheqn} in App.~\ref{app:pseudo} shows a pseudocode for those interested. 

The most substantial difference is that there is no longer any choice of sign within the system itself -- we are instead given a choice for the initial condition $p_{0}$. When there is a choice of branch to take, we will use the notation
\begin{equation}
 	\mathrm{branch}(\cdot) = k ,
\end{equation}
for $k$ labelling the branch, where (as is the case for $p_{0}$) if the choice is of sign we shall use $k = 1$ to signify the positive branch, and $k = 2$ the negative. 



\section{The nonlinear geometry \label{numchap:nls}}

\begin{figure}[htb] \centering
	\includegraphics[width=0.7\textwidth]{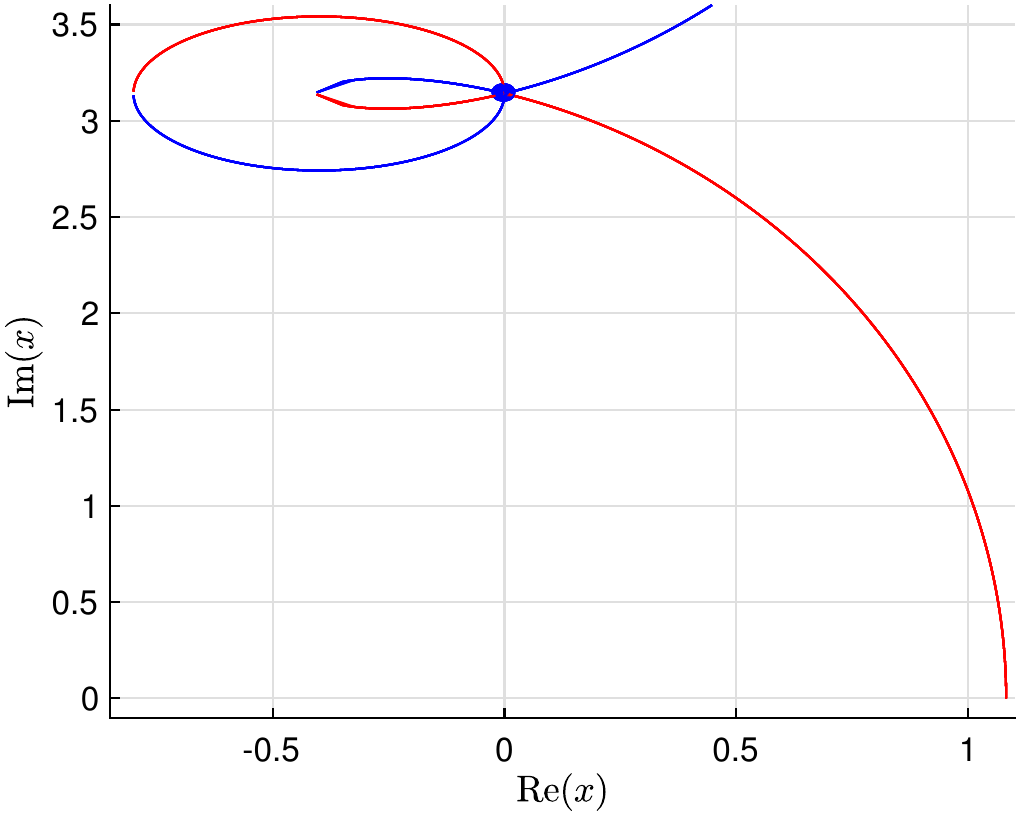}
	\caption{Numerical Stokes lines for the flow \eqref{numnlsphi0}, taking $\rho = 0.01$, $\nu_{j} \in \Theta_{16}$. Red corresponds to $\mathrm{branch}(p_{0}) = 1$, blue to $\mathrm{branch}(p_{0}) = 2$. Again, both lines overlay each other, but this has been prevented for clarity. \label{nlstepnumstokes}}	
\end{figure}

\begin{figure}[htb] \centering
	\includegraphics[width=0.7\textwidth]{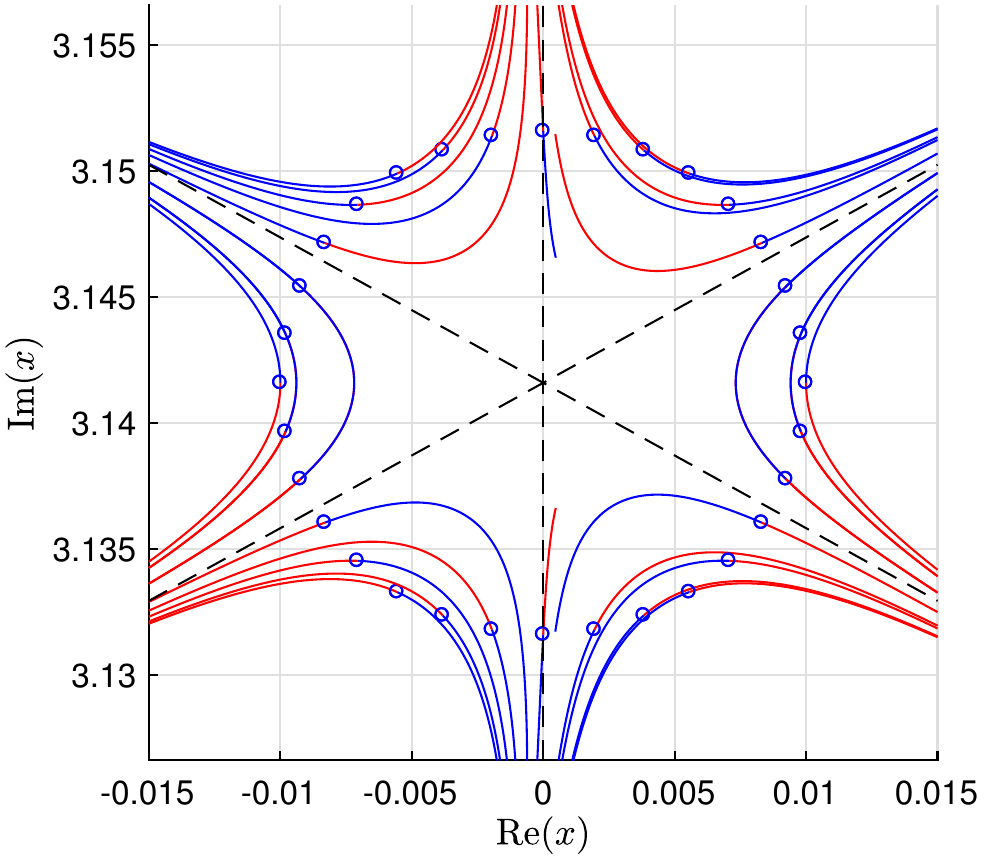}
	\caption{A closer look at the starting points in Fig.~\ref{nlstepnumstokes}, shown as blue circles. Shown dashed black are the equal-phase lines $\theta = (2n - 1)\pi/6, \, n\in\mathbb{Z}$, found from local analysis in Sec.~\ref{nlslocan}. \label{nlstepnumstokeszoomed}}	
\end{figure}

In Sec.~\ref{sec:nls}, we constructed the potential flow given by \eqref{nonlinphi0} with $\delta =1$, that is 
\begin{equation} \label{numnlsphi0}
 	\phi_{0}(x, 0) = x + \frac{1}{2}\log[2(1 + \e^{x})] ,
\end{equation}
and then found the Stokes lines in the complex $x$-plane, shown in Fig.~\ref{stokeslinesnonlin1}. In this section, we will instead use the method in Sec.~\ref{sec:2dcharpm} to find these lines directly, without first solving for $\chi$ analytically. Notice that in general, we only make use of the main branch of \eqref{numnlsphi0} (\emph{cf.} \cite{trinh2013new2}). 
Within the region of interest, all terms in the corresponding Charpit equations are single-valued, and hence we do not need to perform any conversion. The result of now applying the method for the singularity $x = \im\pi$ is shown in Fig.~\ref{nlstepnumstokes}. Comparing this with Fig.~\ref{stokeslinesnonlin1}, we see that once again we have good correspondence. As in Sec.~\ref{numchap:step}, a closer inspection of the region immediately around the singularity (shown in Fig.~\ref{nlstepnumstokeszoomed} with the equal-phase lines found by local analysis in Sec.~\ref{nlslocan}) shows the predicted behaviour.
\vspace{0.3in}

\section{Discussion}
In this chapter we have introduced the important condition \eqref{tautaschitau}, allowing us to numerically follow equal-phase contours. We also supplied a measure, \eqref{fsmeasure}, to evaluate intersection with the real free surface. We then constructed two methods for Stokes lines in two dimensions -- first in Sec.~\ref{sec:2dintm} for problems with $\chi$ formulated as an integral; then in Sec.~\ref{sec:2dcharpm} for the more general situation of characteristic equations. In Sec.~\ref{sec:numbasis}, we also suggested that for functions with branch cuts, it may be necessary to convert to a combination of auxiliary functions to guarantee continuous solutions. We performed such a conversion in Sec.~\ref{numchap:step} for the problem of flow over a step. Finally, we have seen that the line which the method converges to is determined by the sector (bounded by equal-phase lines) within which it begins. 


\chapter{Numerical Stokes surfaces in 3D \label{chap:num3d}}

In this chapter we seek to develop a method that can find Stokes surfaces in 3D. As we will see, new difficulties emerge, the most obvious of which is that we have lost the immediate ability to identify the complexified free surface (now four-dimensional) with real space. To reiterate Sec.~\ref{sec:numbasis}, the one exception to this is the real free surface itself, where $x, \, y \in \mathbb{R}$. 

Unsurprisingly, working in four-dimensional space introduces complications of its own -- for instance our Stokes surfaces are now three-dimensional manifolds, so even visualisation is problematic. As $x$ and $y$ are both complex, it may also be particularly difficult to identify the relevant Riemann sheets. Before continuing to the methods themselves, as the Stokes surface is now formed from the union of all ray solutions over $s \in \mathbb{C}$, we introduce the terminology \emph{Stokes ray} to refer to a single solution for a given $s$. We will consider Stokes rays found to intersect the real free surface to be successful rays. 

\section{The linearised source \label{sec:numlinsource}}
\subsection{The numerical method \label{sec:lin3dnum}}

We first consider the case of the linearised source, with Charpit's equations as in \eqref{linsourcecharp}. The process begins much the same as for those in Chap.~\ref{chap:num2d} -- we apply the chain rule to the ray equations, now with
\begin{equation} \label{lindtaudt}
 	\dd{\tau}{t} = -\overline{p^{4}}/|p^{4}| ,
\end{equation}
to obtain our differential system in $t$, with initial conditions as discussed in Sec.~\ref{sec:3dlin}, and $\tau_{0} = 0$. 
Here, we do not need to worry about bounding $\bm{p}$ as it is constant, and for a given set of initial conditions all terms in \eqref{linsourcecharp} are single-valued so we do not need to perform any conversion. 
To prevent repeating ourselves further, we now only elucidate certain aspects of the method -- Algorithm~\ref{alg:lin3d} in Appendix~\ref{app:pseudo} contains corresponding pseudocode for the full procedure. 
\begin{enumerate}[(i)]
	\item We would predict that analogously to the two-dimensional case, the most important contributions typically come from initial conditions closest to the real free surface -- those with the smallest value of $\mu(\bm{x_{0}}(s))$.  
	\item We must now mesh the complex $s$-plane, that is form a discrete set of $s = s_{r} + \im s_{c}$ for which to evaluate initial conditions. In the case of the linear source we know from \cite{lustri2013steady} that we are aiming for the line $x = 0$ on the real free surface, and additionally our rays are straight. Hence we may predict (or quickly find recursively) an appropriate region of $s$ for a given source depth, $h$. As we will see, for the nonlinear source considered in Sec.~\ref{sec:numnlsource}, this step is not so simple.
	\item By analogy to the Schwarz reflection principle, introduced in Sec.~\ref{sec:nlsstokes}, we expect conjugate initial data to correspond to conjugate terms switching-on (so as to provide real waves on the free surface). 
	
	Suppose that for a given choice of $\branch(y_{0})$ and $\branch(p_{0})$ we have found an $s$ corresponding to a successful ray. To produce conjugate data for $x$ and $y$, from \eqref{pointsourceic} we see that in general we must take $s \mapsto \overline{s}$, and then change $\branch(y_{0})$. In order for the ray to then propagate in the correct direction for intersection, we must additionally change $\branch(p_{0})$. Should we solely take $s \mapsto \overline{s}$ and leave the branches unchanged, we would have $x \mapsto \overline{x}$, $y \mapsto -\overline{y}$ -- this will take the point of intersection $(x, \, y) \mapsto (x, \, -y)$.
	
	For the linear source, recall from Sec.~\ref{sec:3dlin} that we can solve for $x$ and $y$ in terms of $s$ and $\tau$. Combining these solutions with the equation for $\tau$, \eqref{lindtaudt}, we can verify the relations above. 
\end{enumerate}
%

\begin{figure}[htb] \centering
	\includegraphics[width=0.9\textwidth]{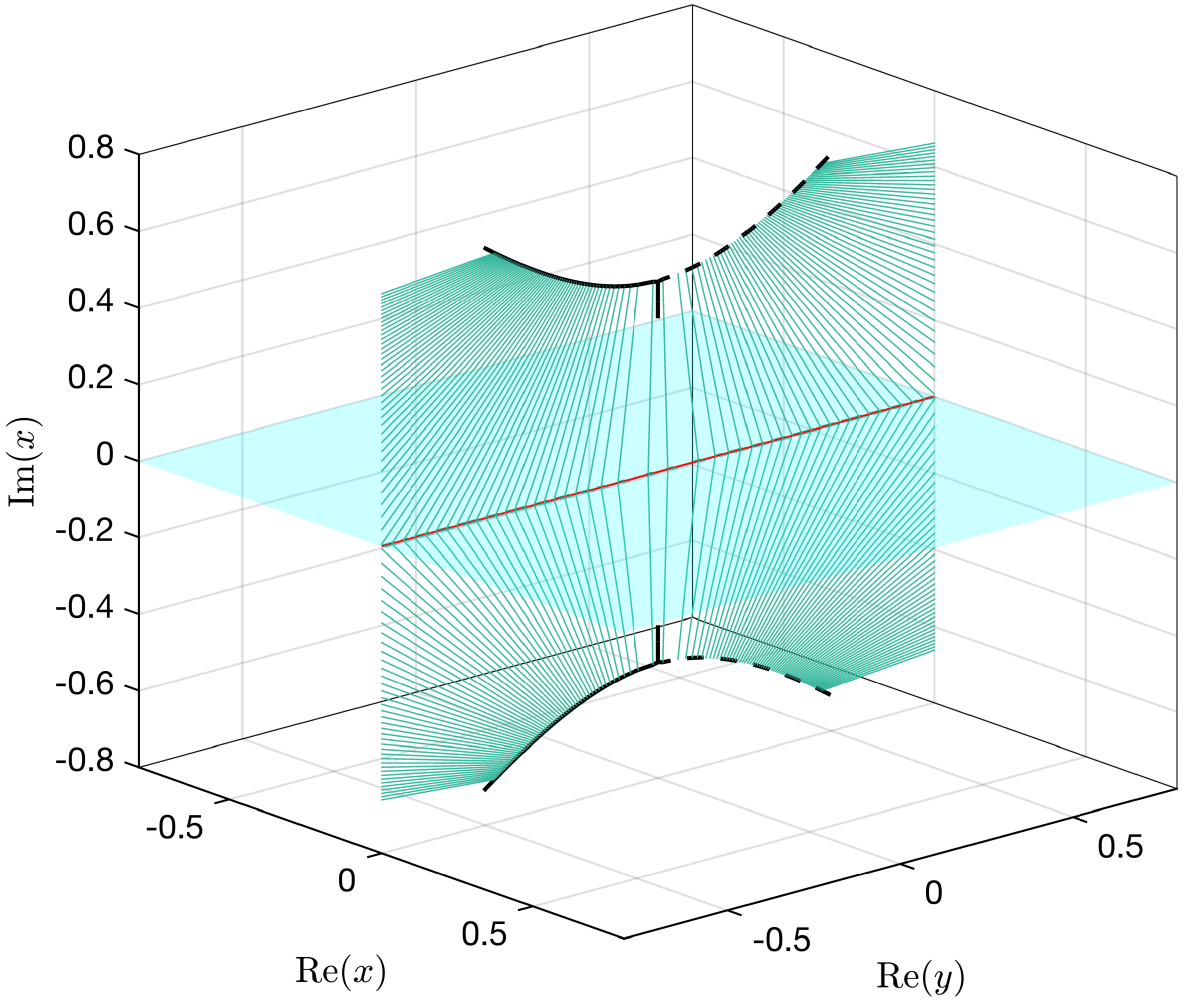}
	\caption{Numerical Stokes rays for the linear source problem with $h = 0.5$. The real free surface (provided $\Im(y)$ is also zero) is shown cyan, and the initial data $x_{0}(s)$, $\Re(y_{0}(s))$ is shown black -- thick for $\mathrm{branch}(y_{0}) = 1$, dashed for $\mathrm{branch}(y_{0}) = 2$. Shown red is the line of intersection found in \cite{lustri2013steady}. Note that the colour of the ray corresponds to the $\Im(y)$ value, but for all intersecting rays this is zero -- hence the lack of variation. 
	\label{linsourcerays}}	
\end{figure}

The result of the method with $h = 0.5$ (chosen for later comparison) is shown in Fig.~\ref{linsourcerays}. The line of intersection found analytically by \cite{lustri2013steady} is shown red, and shows excellent correspondence with the points found by our method.

\subsection{Discussion of the results \label{sec:linsourcedisc}}

Each combination of $\branch(y_{0})$, $\branch(p_{0})$ provides successful rays for one of the four branches of initial data shown. Note that the corresponding initial conditions, $s \in \mathbb{C}\!\setminus\!\mathbb{R}$ with $|s|\ge h$, fall in the exception case above -- $\overline{s}$ does in fact provide conjugate initial data. Here, it is taking the opposite branch for both initial conditions that supplies the point of intersection reflected in $\Re(y) = 0$. In the figure, the rays in $\Re(y) < 0$ correspond to ${(\branch(y_{0}), \, \branch(p_{0})) = (1, \, 2), \, (1, \, 1)}$, and \emph{vice versa} for the rays in $\Re(y) > 0$. As stated, this is the unusual case -- in fact we will see that our conjugacy relations hold for the nonlinear source in Sec.~\ref{sec:numnlsource}.


It should be noted that it is also possible to modify our procedure to instead find the \emph{anti}-Stokes surfaces, where terms associated with the singulant are of equal magnitude to the outer solution -- \emph{i.e.} where $\Re(\chi) = 0$. Analogous logic to Sec.~\ref{sec:numbasis} produces the condition
\begin{equation}
 	\dd{\tau}{t} = \im\overline{\dd{\chi}{\tau}} ,
\end{equation}
for which we may run a comparable method to Sec.~\ref{sec:lin3dnum}. This would provide anti-Stokes lines on the free surface (should they exist), which we could then match to \cite{lustri2013steady},~Fig.~2. For the sake of brevity we will not do so here.

\section{The nonlinear source 	\label{sec:numnlsource}}
\subsection{The numerical method}

Once more, the method for the nonlinear source follows the same initial procedure -- we apply the chain rule to the corresponding Charpit's equations, \eqref{neatnlcheqns}, with 
%
%
\begin{equation} \label{dtaudt}
 	\dd{\tau}{t} = -\overline{(\bm{p}\cdot\bm{p})}/|\bm{p}\cdot\bm{p}| ,
\end{equation}
and initial conditions as discussed in Sec.~\ref{sec:initsource}, again with $\tau_{0} = 0$.  We will bound $|\bm{x}|$ and $|\bm{p}|$, and include the equation for $\phi_{0}$ using \eqref{phi0teqn}. As the derivatives of $\phi_{0}$ contain branch cuts, as performed in Sec.~\ref{numchap:step} we should convert to a combination of auxiliary functions, and then include the corresponding equations. However, for the problem considered we obtained better results without introducing the additional error from doing so, and simply verified afterwards that all of our solutions were continuous. For more general velocity potentials it may be necessary to perform the conversion process. 

The nonlinear method faces additional difficulties to the procedure in Sec.~\ref{sec:lin3dnum}. We illuminate the key differences here (see Algorithm~\ref{alg:nl3d} in App.~\ref{app:pseudo} for the relevant pseudocode). 
\begin{enumerate}[(i)]
	\item When setting initial conditions we are now starting off the singularity, and importantly must choose values of $\nu, \, \vartheta$ in \eqref{nlsourceic}. Analogous to Chap.~\ref{chap:num2d}, we expect there to be four-dimensional sectors (bounded by 3D equal-phase surfaces) which determine the Stokes surface our method converges to. As we do not predict all possible Stokes surfaces to intersect with the real free surface, we must hence vary these angles. In practice it is too computationally expensive to do so for a large number of values -- for our numerics we will take $\nu, \, \vartheta \in \{0, \, \pi\}$, as this turns out to be sufficient to find successful rays. 
	\item As we will see, intersection is highly sensitive to the initial conditions provided. Consequently, we can no longer rely on being able to produce successful rays simply by meshing regions of $s \in \mathbb{C}$.
	\item The key step is to reformulate the system in a way amenable to the shooting method (see \emph{e.g.} \cite{press1989numerical}). First, we consider the value of $\mu(\bm{x})$ at the ray endpoint to be a function of $s$, and coarsely mesh a large region of complex $s$-space to supply initial guesses, $s_{0} \in S_{0}$. This allows us to use a numerical nonlinear system solver (\emph{e.g.} fsolve in \textsc{Matlab}) to try and find roots of the function, where $\mu(\bm{x}_{\text{end}};\,s) = 0$. Now, rather than expecting $\mu(\bm{x})=0$, 
	to allow for high sensitivity we instead impose the relaxed condition that numerical integration stop when either $\Im(x)$ or $\Im(y)=0$. This step is necessary to 
	ensure that the solver converge, else we do not obtain sufficiently small $\mu(\bm{x}_{\text{end}};\,s_{0})$.
	\item The method is now recursive -- for $s_{0} \in S_{0}$ that allow convergence, we reduce the mesh to around these values (to a new mesh, $S_{1}$), and repeat the shooting method. We then iterate this process until a desired number of intersecting rays have been found. We must still check that these rays satisfy the additional condition $\Im(\phi_{0}) = 0$ upon intersection. 
\end{enumerate}
%

\begin{landscape}
\begin{figure}[htb] \centering
	\includegraphics[width=0.75\linewidth]{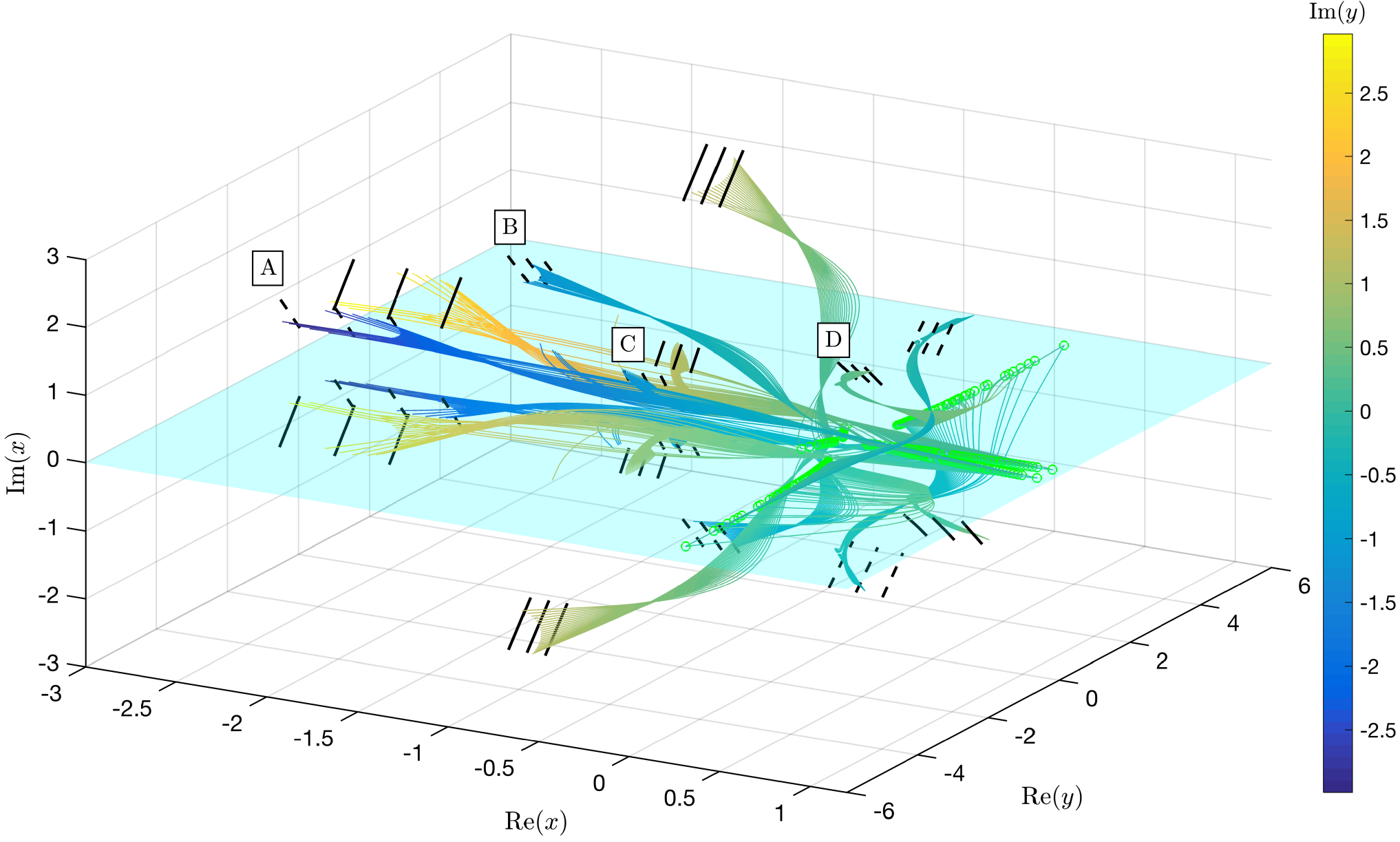}
	\caption{Numerical Stokes rays for the nonlinear source problem. The real free surface is shown cyan, $(x_{0}(s_{0})$, $\Im(y_{0}(s_{0})))$ 
	shown black -- thick for $\branch(y_{0}) = 1$, dashed for $\branch(y_{0}) = 2$. Endpoints of the rays are highlighted green (see Fig.~\ref{nlsourcehits}). We have taken $h = 0.5$, $\delta = -0.3$, $\rho = 0.015$. Each set of three black lines corresponds to the final mesh region, $S$, for one choice of $\branch(y_{0})$, $\branch(p_{0})$, $\nu$ and $\vartheta$. Labels are discussed in Sec.~\ref{sec:nlsourcedisc}.
	\label{nlsourcerays}}	
\end{figure}
\end{landscape}

The Stokes rays resulting from applying this procedure with $h = 0.5$, $\delta = -0.3$, and $\rho = 0.015$ are shown in Fig.~\ref{nlsourcerays}, with corresponding points of intersection in Fig.~\ref{nlsourcehits}. Note that some points overlay each other, and so do not appear on this figure -- Figs.~\ref{nloppbranches}~and~\ref{nlsamebranches} clarify this. Unlike source depth in the linearised problem, changing $h$ and $\delta$ does indeed have an effect on the line of intersection. We have chosen the given values relatively arbitrarily, but it should be noted that increasing $h$ increases the distance successful rays must travel (and so the method converges more slowly), and $\delta$ affects the curvature of the rays, which predictably can have a large impact on intersection locations. 

\begin{figure}[htb] \centering 
	\includegraphics[width=0.9\textwidth]{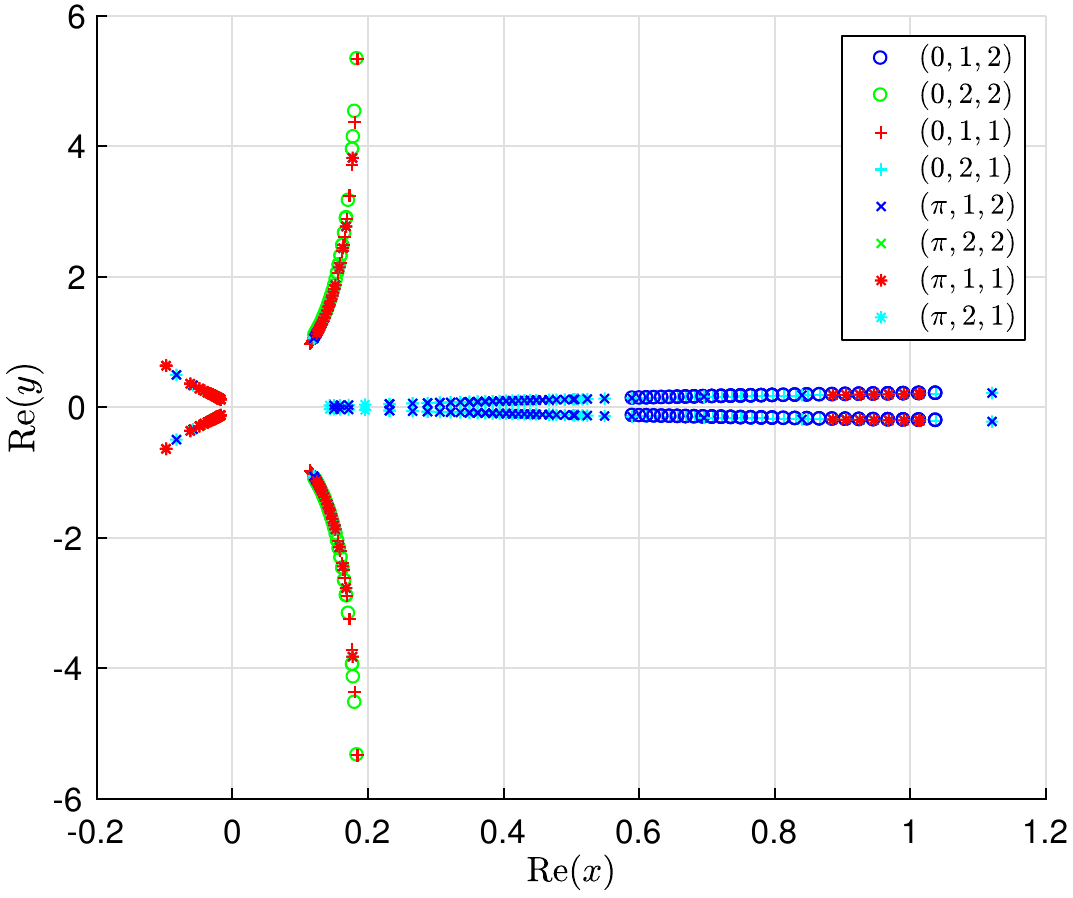}
	\caption{Points of intersection of Stokes rays with the real free surface for the nonlinear source problem. We have taken $h = 0.5$, $\delta = -0.3$, $\rho = 0.015$. The legend shows corresponding $(\nu, \, \mathrm{branch}(y_{0}), \, \mathrm{branch}(p_{0}))$ values. 
	\label{nlsourcehits}}	
\end{figure}
\begin{figure}[p] \centering \ContinuedFloat
	\includegraphics[width=0.8\textwidth]{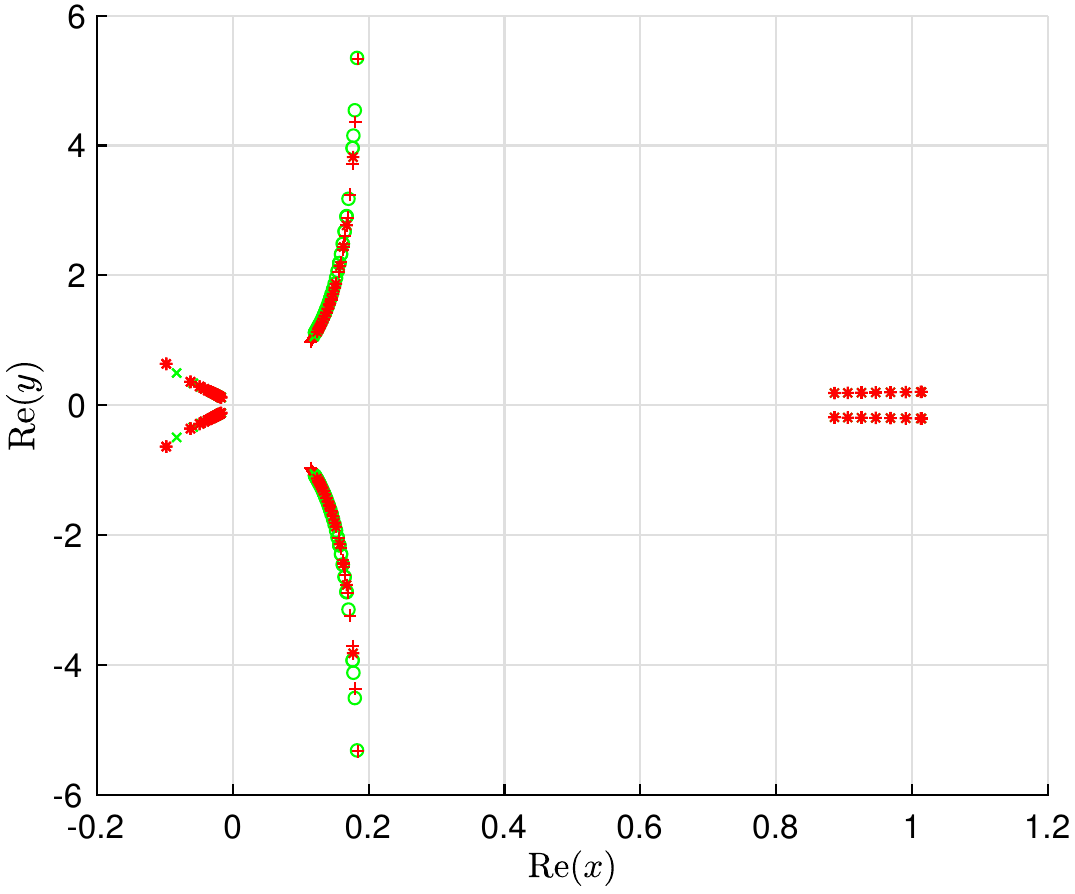}
	\caption{Points of intersection for Stokes rays with initial conditions in case (i). \label{nlsamebranches}}
\end{figure}
\begin{figure}[p] \centering \ContinuedFloat
	\includegraphics[width=0.8\textwidth]{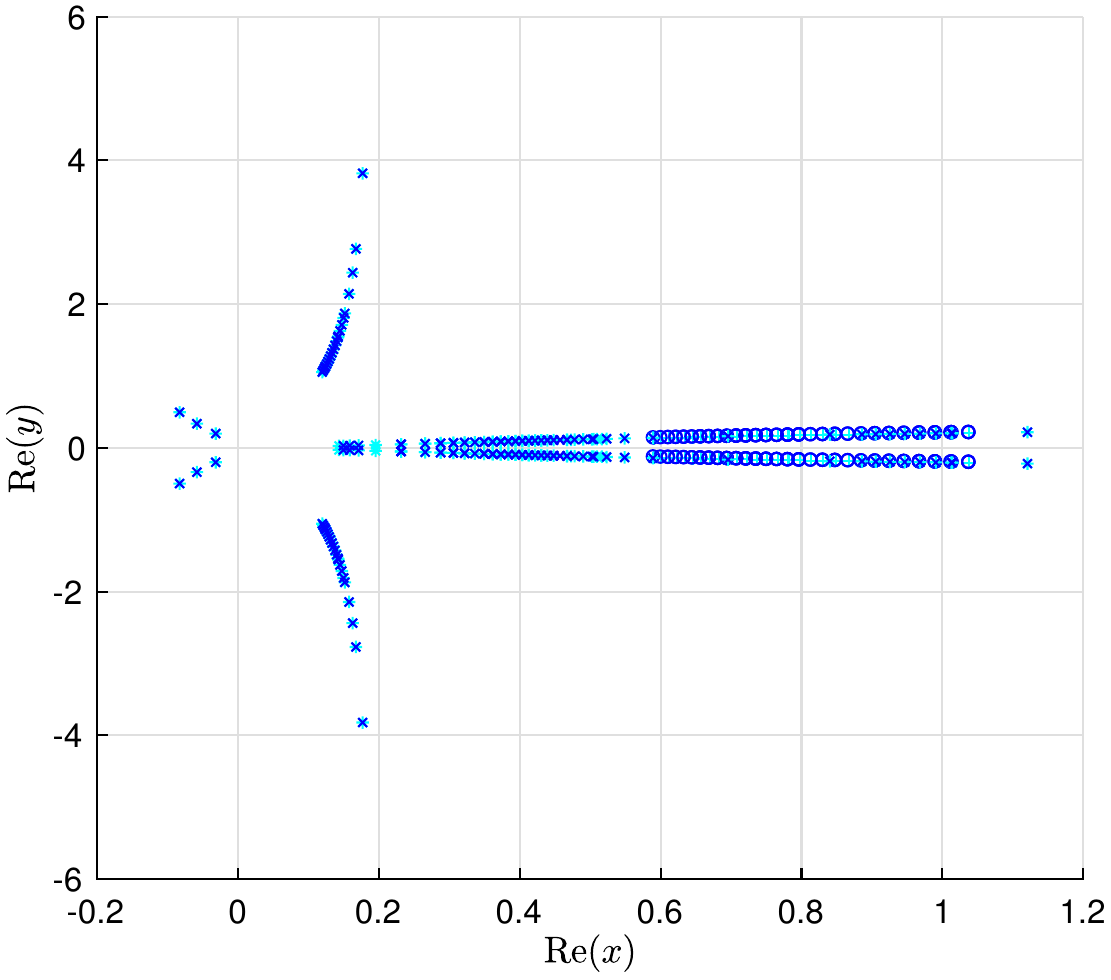}
	\caption{Points of intersection for Stokes rays with initial conditions in case (ii). \label{nloppbranches}}
\end{figure}

\begin{figure}[htb] \centering
	\includegraphics[width=0.9\textwidth]{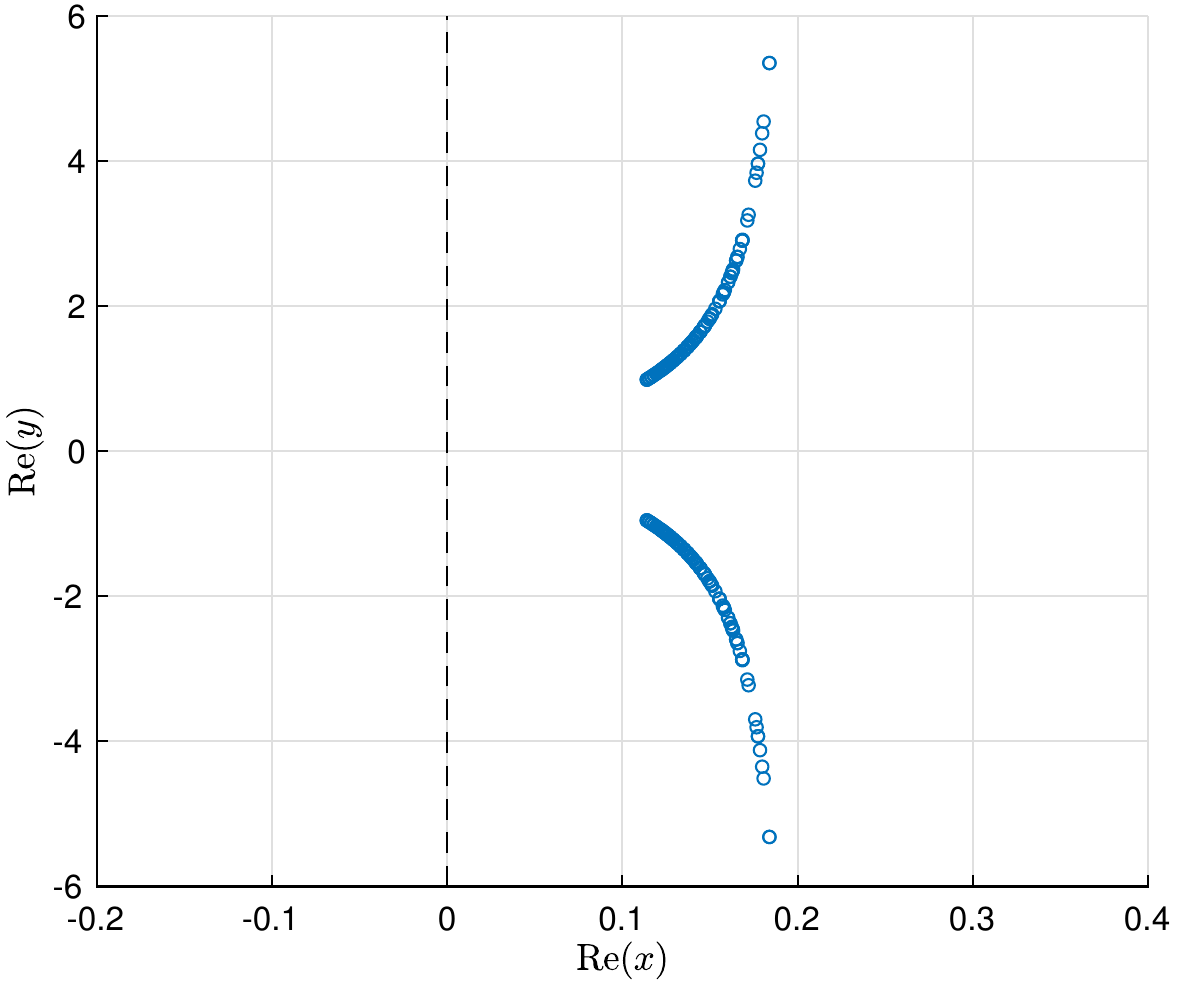}
	\caption{Shown blue are the points of intersection for $\nu = 0$ and $\branch(y_{0}), \, \branch(p_{0})$ of the same sign. Shown dashed black is the Stokes line predicted from the linear method in \cite{lustri2013steady}. \label{chap5:wings}}
\end{figure}

\subsection{Discussion of the results \label{sec:nlsourcedisc}}

The initial data for the problem satisfies the same conjugacy relations upon $s \mapsto \overline{s}$ as those described in Sec.~\ref{sec:lin3dnum}, so long as we also change the choice of $\vartheta$. As such, we would predict the same behaviour for the rays, that is $(x,\,y) \mapsto (\overline{x},\, -\overline{y})$ if we keep the same branch choices, and $(x,\,y) \mapsto (\overline{x},\, \overline{y})$ if we switch both. This means we have two distinct cases: (i) $\branch(y_{0}), \, \branch(p_{0})$ take the same sign, or (ii) $\branch(y_{0}), \, \branch(p_{0})$ take opposite signs. 

We also have the choice of $\nu \in \{0, \, \pi\}$. In Fig.~\ref{nlsourcerays}, we mark the upper left branch of each combination -- A corresponds to case (ii) with $\nu = \pi$, B to (i) with $\nu = \pi$, C to (ii) with $\nu = 0$, and finally D to (i) with $\nu = 0$. 
We observe that the choice of $\nu$ changes the necessary region of $s$. For the two cases (i) and (ii) to be the crucial determinant of points of intersection, we require this to have little impact. From Fig.~\ref{nlsourcehits}, we see that indeed this is indeed the case. 

As we can see from Figs.~\ref{nloppbranches}~and~\ref{nlsamebranches}, both cases result in the same structure of intersecting points, but with differing distributions. As we have previously stated, the system is highly sensitive to initial conditions; we can see from Fig.~\ref{nlsourcerays} that a simple meshing procedure (shooting from the black lines) would have failed to pick up almost all the rays, even had we somehow ascertained that these were the optimal regions to mesh in the first place. Additionally, sensitivity lies within the system itself -- these figures were produced with the smallest feasible tolerances for both the numerical integration and the numerical nonlinear system solver. As we decreased these tolerances the distribution of points of intersection changed, typically moving from the narrow wedge structure to the leftmost `spikes', and then finally to the `wings'.
This remains an issue for further work, but we emphasise that our key result is the general qualitative structure of the Stokes surface intersections, rather than the precise intersection values. 

Ideally, we would like to be able to discount the wedge and spike structures on the basis that the analogous 3D linear problem of \cite{lustri2013steady} as well as the nonlinear 2D problem of \cite{chapman2006exponential} do not seem to exhibit the same phenomena. 

We found that for rays starting in case D of Fig.~\ref{nlsourcerays}, these structures did not appear -- only the wings remained. The points of intersection are shown in Fig.~\ref{chap5:wings} along with the line predicted by \cite{lustri2013steady}. These are the rays which travel the shortest distance before intersecting the real free surface. Combining this with the changing behaviour of the system under increasingly strict tolerance, we conjecture that the spikes and wedge are thus the products of numerical error. To narrow the gap inbetween the wings, we must provide initial conditions, $s$, of increasing magnitude -- Fig.~\ref{addfig:nlwings2} in Appendix~\ref{app:addfigs} provides a visualisation of this.

\section{Discussion}
In this chapter we have produced a numerical method capable of finding Stokes rays for the nonlinear source problem introduced in Chap.~\ref{chap:expascrt}. Note that the method itself is applicable to far more general potentials, and the case of the source is useful due to its possible use as a Green's function to model three-dimensional obstructions. We also developed a simpler procedure -- which did not require reformulating the system in a way amenable to the shooting method -- for the linearised source problem of \cite{lustri2013steady}, and recovered the same free surface Stokes line. 

%
%
\chapter{Conclusions and future work \label{conclusion}} 

The aim of this dissertation was to develop and apply a procedure in order to find the Stokes surfaces connected with the study of steady low-Froude flows in three dimensions. These Stokes surfaces lie in four-dimensional space, and are connected with singularities imposed by the geometry of the wave-making body. They are specified in connection with a complex-ray formulation of Laplace's equation. The intersection of these Stokes surfaces with the physical free surface indicates a critical location whereby surface waves are switched-on by the Stokes Phenomenon.

Our work serves as an extension to the investigation of \cite{lustri2013steady}, where they had derived the Stokes-surface intersections in the case of flow past a linearised source. In this linear-geometrical context, the intersection can be predicted by solving Charpit's equation analytically. In the case of nonlinear geometries, no such analytical solution is possible in general. Thus our method seeks to develop the necessary numerical procedures for resolving the Stokes surfaces. 

Our key result, shown in Fig.~\ref{chap5:wings} confirms that for the case of nonlinear geometries, the Stokes surfaces are curved manifolds. This result is expected based on the situation of flows past bodies in two-dimensions---here, linear geometries produce straight-line Stokes lines, whereas nonlinear geometries product curved Stokes `lines'. We believe that this is the first visualisation of a Stokes surface for the nonlinear problem. 




There are many exciting avenues for further study. We will outline some possibilities here:
\begin{enumerate}[(i)]
\item One substantial area of interest is how the Stokes surfaces can be projected into the physical fluid. As visualised in Fig.~\ref{3dstokes}, we expect that Stokes surfaces emanate through the fluid from critical points of three-dimensional obstructions, yet our current method can only find their intersection with the free surface. That such a projection must exist is sensible due to physical reasons. After all, if waves switch-on at the free surface, then they must also switch-on slightly below the free-surface. This switching-on region must tend to zero as the Froude number tends to zero, and hence it would be expected that our Stokes surfaces can be projected into $(x,y,z)$-space. For the two-dimensional problem, this projection emerges as a consequence of the uniqueness of analytic continuation (cf. Fig.~6 of \cite{trinh_tulinmodel}).

\item As seen in Sec.~\ref{sec:numnlsource}, determining necessary initial conditions in 4D space was difficult -- a local analysis in characteristic variables would be the first step to resolve this, by better understanding of the local topology of the Stokes surfaces. \label{icfurtherwork}

\item Throughout this dissertation, we have neglected surface tension -- the development of a numerical procedure that could produce the Stokes surfaces for gravity-capillary waves would be a powerful extension. We show how to begin the process in Appendix~\ref{app:surfacetension}. 

\item Within our method for the nonlinear source in Sec.~\ref{sec:numnlsource}, we relaxed the stopping condition for the integrator from $\mu(\bm{x}) = 0$, to $\Im(x)$ or $\Im(y) = 0$. While this was necessary to obtain convergence, it also removes the possibility of rays curving across one imaginary axis and then hitting the free-surface. The resolution of this is connected to \eqref{icfurtherwork}. 

\item In analytically solving for $\chi$, \cite{lustri2013steady} find eight possible singulant expressions, with two distinct behaviours that are physically relevant corresponding to either longitudinal or transverse waves. We believe this is connected to the two cases of branch combinations seen in Chap.~\ref{chap:num3d}. The resolution of this conjecture is linked to another problem: how can $\chi$ be determined elsewhere on the free surface? To do so would be necessary in order to establish subdominant wave-switching locations---these are where different singulants interact with each other via the \cite{dingle1973asymptotic} conditions \eqref{eq:dingle}.

\end{enumerate}


\begin{appendices}

\chapter{An extension to include surface tension \label{app:surfacetension}}

Due to the apparent success of our nonlinear numerical method in Sec.~\ref{sec:numnlsource}, further work could pursue an analogous process for the Stokes surfaces of free-surface gravity-capillary waves. In the hope of stimulating such study, we begin the process here. 

If we now include surface tension, the nondimensional Bernouilli's equation on the free surface $z=\eta$ becomes
\begin{equation} \label{beqnst}
 	\frac{F^{2}}{2}\left(|\nabla\phi|^{2} - 1\right) + z = -B\kappa , 
\end{equation}
where $\kappa$ is the mean curvature of the surface,
\begin{equation}
 	\frac{(1+\eta_{y}^{2})\eta_{xx} + (1+\eta_{x}^{2})\eta_{yy} - 2\eta_{x}\eta_{y}\eta_{xy}}{(1+\eta_{x}^{2}+\eta_{y}^{2})^{3/2}} ,
\end{equation}
and $B$ is the Bond number (sometimes called the inverse of this)
\begin{equation}
 	B = \frac{\sigma}{\rho g L^{2}} ,
\end{equation}
with $\sigma$ the surface tension parameter, $\rho$ the density of the fluid, and once more $L$ a typical length-scale. As before, $F$ denotes the Froude number given by \eqref{froudenumber}. 

As in \cite{trinh2013new}, we are interested in the low-Froude, low-Bond limit. As such, we let
\begin{equation}
 	F^{2} = \beta\ep, \quad B = \beta\tau\ep^{2} ,
\end{equation}
where $\ep\ll 1$. This choice of $F$, $B$ is so that after performing our late-order analysis, both sides of equation \eqref{beqnst} feature. Substituting in, we now have
\begin{equation} \label{beqnst1}
 	\frac{\beta\ep}{2}\left(|\nabla\phi|^{2} - 1\right) + z = -\beta\tau\ep^{2}\kappa .
\end{equation}
\vspace{-0.1in}
Applying the change of variable $Z = z - \eta$, the mean curvature becomes
\begin{multline}
 	\kappa = \biggl[(1+\eta_{x}^{2}(1-\eta_{Z})^{2})(\eta_{xx}(1-\eta_{Z}) - 2\eta_{x}\eta_{xZ} + \eta_{x}^{2}\eta_{ZZ}) \\
	+ (1+\eta_{y}^{2}(1-\eta_{Z})^{2})(\eta_{yy}(1-\eta_{Z}) - 2\eta_{y}\eta_{yZ} + \eta_{y}^{2}\eta_{ZZ}) \\
	- 2(\eta_{x}\eta_{y}(1-\eta_{Z})^{2})(\eta_{xy}(1-2\eta_{Z}) - \eta_{y}\eta_{xZ} - \eta_{x}\eta_{yZ} + \eta_{x}\eta_{y}\eta_{ZZ})\biggr] \\
	/ (1 + (\eta_{x}^{2}+\eta_{y}^{2})(1-\eta_{Z})^{2})^{3/2} .
\end{multline}
While this expression seems particularly unpleasant, we may use the same logic as in Chap.~\ref{chap:expascrt} to only consider the dominant terms. That is if we again expand $\eta$ and  $\Phi$ (where recall $\Phi(x,\,y,\,Z) = \phi(x,\,y,\,z)$) as power series in $\ep$, \eqref{beqnst1} still provides $\eta_{0}=0$, and hence we find that at $\Oh(\ep^{n})$ as $n \to \infty$ the leading-order mean curvature is 
\begin{equation}
	\kappa \sim \eta_{(n - 2) xx} + \eta_{(n - 2) yy}.
\end{equation}

Now substituting our factorial-over-power ansatz (as in \eqref{3dloansatz}) into \eqref{beqnst1}, we consequently find the corresponding leading-order expression for late-order terms to be
\begin{equation} \label{lodbc2}
 	-\beta A(\chi_{x}\Phi_{0 x} + \chi_{y}\Phi_{0 y}) + B = -\beta\tau B(\chi_{x}^{2}+\chi_{y}^{2}) .
\end{equation}
The kinematic boundary condition on the free surface remains the same as in Chap.~\ref{chap:expascrt}, providing \eqref{lokbc1}, as does Laplace's equation within the fluid, providing the eikonal equation \eqref{eikonal3d1}. Hence we may solve \eqref{lodbc2} simultaneously with \eqref{lokbc1} to find that for non-trivial $\chi$ on the free surface $Z = 0$, we require	
\begin{equation}
 	\chi_{Z} = \frac{\beta(\chi_{x}\Phi_{0 x}+\chi_{y}\Phi_{0 y})^{2}}{(1+\beta\tau(\chi_{x}^{2}+\chi_{y}^{2}))} .
\end{equation}
Substituting this into \eqref{eikonal3d1} we thus find
\begin{equation}
 	\chi_{x}^{2} + \chi_{y}^{2} + \frac{\beta^{2}(\chi_{x}\Phi_{0 x}+\chi_{y}\Phi_{0 y})^{4}}{(1+\beta\tau(\chi_{x}^{2}+\chi_{y}^{2}))^{2}} = 0 .
\end{equation}
By the same argument as in Chap.~\ref{chap:expascrt}, to leading order we may replace $\Phi$ with $\phi$, and hence obtain the necessary equation to continue as before. Also note that if we neglect surface tension in this equation so that $\tau = 0$, and set $\beta = 1$ (by choice of scaling), we recover the previous formula \eqref{3dnonlineikonal_again}. The situation of gravity-capillary waves is more involved than that dealt with in this dissertation -- for instance Stokes lines/surfaces may emanate from stagnation points as well as singularities \citep{trinh2013new2}. We leave the interesting possibility of a nonlinear numerical procedure to further study.

\chapter{On numerics}

\section{Additional figures \label{app:addfigs}}
\begin{figure}[htb] \centering
	\includegraphics[width=1.1\textwidth]{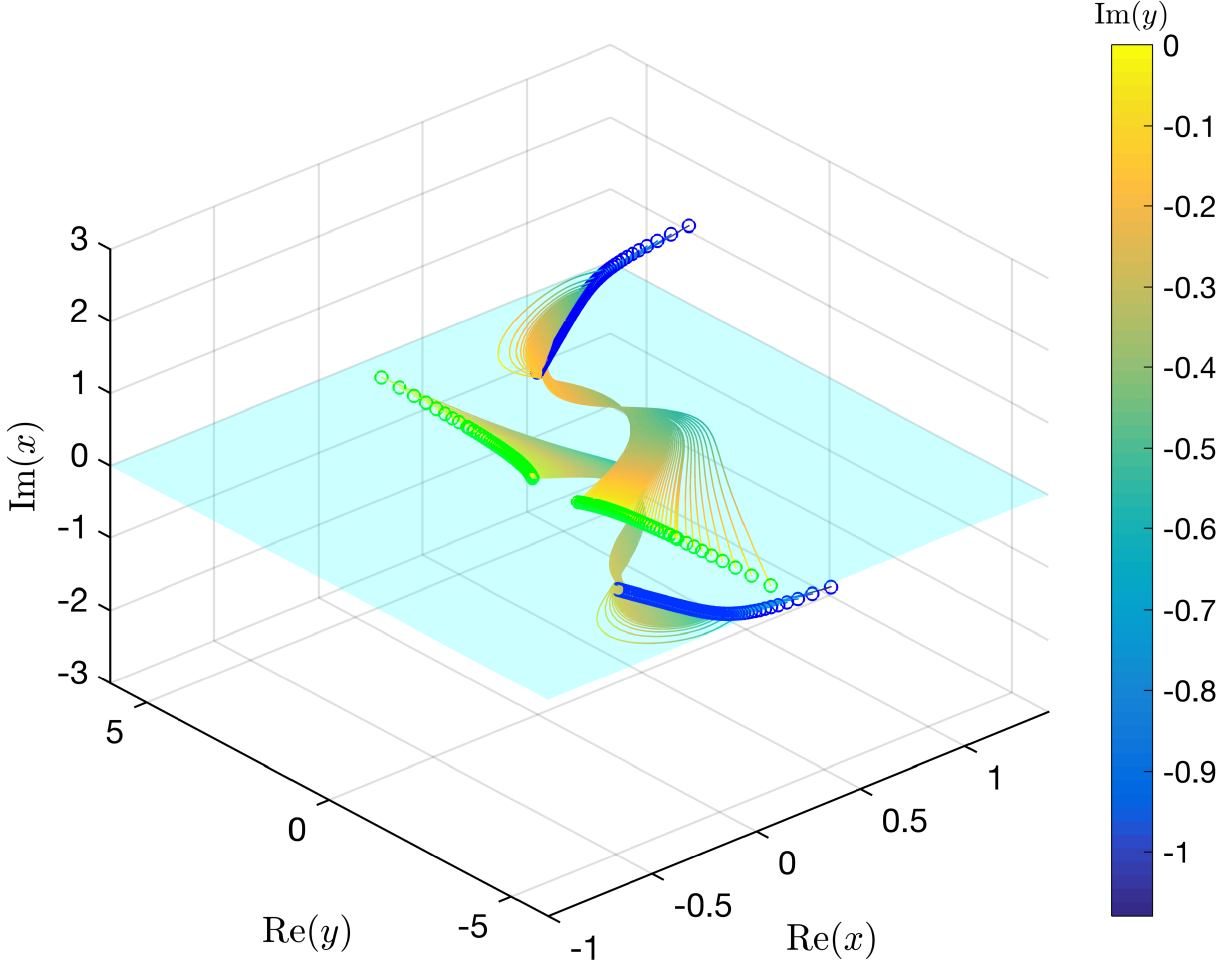}
	\caption{A subset of the Stokes rays from Fig.~\ref{nlsourcerays} -- here $\nu = 0$, $\branch(y_{0}) = \branch(p_{0}) = 2$. The real free surface is shown cyan. Blue circles show the starting point of a ray, green circles show a point of intersection.\label{addfig:nlwings2}}
\end{figure}

\begin{figure}[htb] \centering
	\includegraphics[width=1\textwidth]{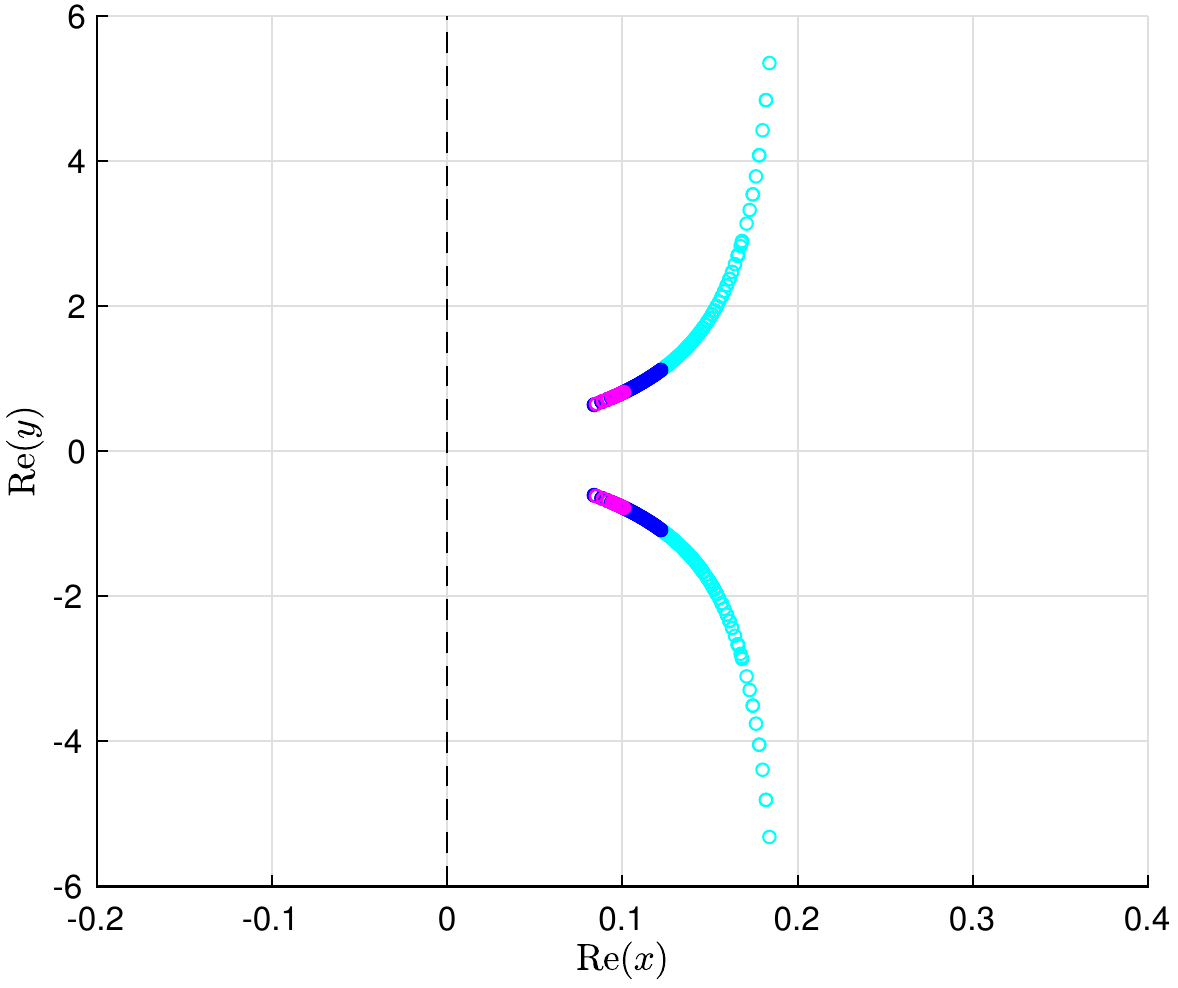}
	\caption{Points of intersection for $\nu = 0$, $\branch(y_{0}) =  \branch(p_{0}) = 2$. The different colours correspond to different mesh regions taken in the method of Sec.~\ref{sec:numnlsource}. As the curve tends towards $\Re(x) = 0$ the magnitude of the necessary initial condition, $s$, must increase -- cyan points have smaller $|s|$ than blue points, which have smaller $|s|$ than magenta points. Shown dashed black is the Stokes line predicted from the linear method in \cite{lustri2013steady}. \label{addfig:nlwings2}}
\end{figure}

\begin{landscape}
\section{Summary of methods \label{app:pseudo}}
\begin{minipage}[t]{0.4\linewidth}
\begin{algorithm}[H]
\caption{2D method via integral} \label{alg:2dint}
\begin{algorithmic}[1]
	\State Choose singularity
	\State If necessary, convert terms as discussed, then set derivatives of variables according to \eqref{numintxt} and \eqref{numintcpt}
	\State Choose a spacing, $\rho$, number of divisions of $\pi$, $n$, sign in \eqref{numdchidx}, bounds for $|x|$, $|\chi'(x)|$ and desired tolerance for $\Im(\phi_{0})$
	\For{$\theta_{j}$} 
		\State Set initial conditions
		\State Numerically integrate for negative $t$
		\If{$\Im(x) = 0$ \textbf{and} $\Im(\phi_{0}) <$ tolerance at endpoint}
			\State Stop integration, plot $x$
		\ElsIf{$|x|$ or $|\chi'(x)|$ exceed chosen bounds}
			\State Stop integration
		\EndIf
	\EndFor	  
\end{algorithmic}
\end{algorithm}
\end{minipage}
\hspace{0.5in}
\begin{minipage}[t]{0.4\linewidth}
\begin{algorithm}[H]
\caption{Nonlinear 2D method via Charpit's equations} \label{alg:nl2dcheqn}
\begin{algorithmic}[1]
	\State Choose singularity
	\State Set derivatives of variables according to \eqref{numchap:2dcharpitseqns}, with the Charpit's equations given by \eqref{2dcharpiteqns}, $\tau'$ by \eqref{2ddtaudt} and corresponding $\phi_{0}'$
	\State Choose a spacing, $\rho$, number of divisions of $\pi$, $n$, bounds for $|x|$ and $|p|$ and desired tolerance for $\Im(\phi_{0})$
	\For{$\mathrm{branch}(p_{0}) = 1, \, 2$}
    		\For{$\theta_{j}$} 
    			\State Set initial conditions
	    		\State Numerically integrate for negative $t$
    			\If{$\Im(x) = 0$ \textbf{and} $\Im(\phi_{0}) <$ tolerance at endpoint}
    				\State Stop integration, plot $x$
			\ElsIf{$|x|$ or $|p|$ exceed chosen bounds}
				\State Stop integration
	    		\EndIf
    		\EndFor	 
	\EndFor
\end{algorithmic}
\end{algorithm}
\end{minipage}
\hspace{-0.35in}
\begin{minipage}[t]{0.45\linewidth}
\vspace{-1.2in}
\begin{algorithm}[H]
\caption{3D linearised source method} \label{alg:lin3d}
\begin{algorithmic}[1]
	\State Set depth of source, $h$, bound for $|\bm{x}|$, and desired tolerance for $\mu(\bm{x})$
	\State Set derivatives of variables by applying the chain rule to the Charpit's equations given by \eqref{linsourcecharp}, with $\tau'$ given by \eqref{lindtaudt}
	\State Form a discrete mesh, $S$, of a region of the complex $s$-plane
	\For{$s \in S$}
		\For{$\mathrm{branch}(y_{0}) = 1, \, 2$}
			\For{$\mathrm{branch}(p_{0}) = 1, \, 2$}
                    		\State Set initial conditions
                    		\State Numerically integrate for negative $t$
                        		\If{$\mu(\bm{x}) <$ tolerance}
                    			\State Stop integration, plot ray
				\ElsIf{$|\bm{x}|$ exceeds chosen bound}
					\State Stop integration
                    		\EndIf
			\EndFor
		\EndFor
	\EndFor
\end{algorithmic}
\end{algorithm}

\end{minipage}
\hspace{0.4in}
\begin{minipage}[t]{0.45\linewidth}
\vspace{-1.2in}
\begin{algorithm}[H]
\caption{3D nonlinear source method} \label{alg:nl3d}
\begin{algorithmic}[1]
	\State Set depth of source, $h$, source strength, $\delta$, spacing $\rho$, bounds for $|\bm{x}|$ and $|\bm{p}|$, and tolerances for numerical solvers and $\Im(\phi_{0})$
	\State Set derivatives of variables by applying the chain rule to the Charpit's equations given by \eqref{neatnlcheqns}, with $\tau'$ given by \eqref{dtaudt}
	\State Form a coarse discrete mesh, $S$, of a large region of the complex $s$-plane
	\For{$s_{0} \in S$}
		\For{$\mathrm{branch}(y_{0}), \, \mathrm{branch}(p_{0})  \in \{1, \, 2\}$}
			\For{$\theta, \, \vartheta \in \{0, \, \pi\}$}
				\For{numerical nonlinear system solver, as a function of $s$ \label{nl3dshootloop}}
                    			\State Set initial conditions
                    			\State Numerically integrate for negative $t$
                        			\If{$\Im(x)$ or $\Im(y) = 0$}
                    				\State Stop integration, store solution details
					\ElsIf{$|\bm{x}|$ \textbf{or} $|\bm{p}|$ exceed chosen bounds}
						\State Stop integration
                    			\EndIf
				\EndFor
				\State Reduce mesh region $S$ to around successful shooting points
				\State Repeat \ref{nl3dshootloop} until desired number of successful $s$ points found
				\If{$\Im(\phi_{0}) <$ tolerance at endpoint}
					\State Plot ray
				\EndIf
			\EndFor
		\EndFor
	\EndFor
\end{algorithmic}
\end{algorithm}
\end{minipage}

\end{landscape}

\end{appendices}

\backmatter


\bibliographystyle{plainnat}
\bibliography{JF-CCD-bib}        

\end{document}